\documentclass[10pt, letterpaper, notitlepage, oneside]{article}

\usepackage[
	letterpaper,
	bottom      = 1.00in,
	left        = 0.75in,
	right       = 0.75in,
	top         = 1.00in
]{geometry}
\usepackage[utf8]{inputenc}
\usepackage[nopar]{lipsum}

\usepackage{fancyhdr}
\fancypagestyle{specialfooter}{
	\fancyhf{}
	\lhead{Published in \textit{Physics of Fluids}: \href{https://doi.org/10.1063/1.5083054}{\url{https://doi.org/10.1063/1.5083054}} \hfill \thepage}
	\cfoot{$\dag$ \textbf{Email address for correspondence:} g\_dilabb{\MVAt}encs.concordia.ca}
}

\usepackage{hyperref}
\hypersetup{
	breaklinks  = true,
	citecolor   = blue,
	colorlinks  = true,
	anchorcolor = blue,
	filecolor   = blue,
	linkcolor   = blue,      
	urlcolor    = blue
}

\usepackage{titlesec}
\titleformat*{\section}{\Large\bfseries}
\titleformat*{\subsection}{\normalsize\bfseries\filcenter}
\titleformat*{\subsubsection}{\normalsize\bfseries}
\renewcommand{\thesection}{\Roman{section}}

\makeatletter
\renewcommand\@seccntformat[1]{\csname the#1\endcsname.\quad}
\makeatother

\usepackage{amsfonts}
\usepackage{amsmath}
\usepackage{marvosym}
\usepackage{textcomp}
\usepackage{upgreek}
\usepackage[x11names,rgb]{xcolor}
\newcommand{\volume}{{\ooalign{\hfil$V$\hfil\cr\kern0.08em--\hfil\cr}}}

\usepackage{caption}
\usepackage{cleveref}
\usepackage{graphicx}
\usepackage{epstopdf}
\usepackage[
	font = normalsize
]{subfig}
\usepackage{xparse}
\crefformat{subfigure}{#2\onlyletter{#1}#3}
\crefrangeformat{subfigure}{#1--#3\onlyletter{#2}#4}
\ExplSyntaxOn
\NewDocumentCommand{\onlyletter}{m}{
	\tl_set:Nx  \l_tmpa_tl { #1 }
	\tl_item:Nn \l_tmpa_tl { -1 }
}
\ExplSyntaxOff

\usepackage{arydshln} 
\usepackage{multirow}
\usepackage{setspace}

\usepackage{apacite} 
\let\cite\shortcite
\let\citeA\shortciteA

\newcommand{\noop}[1]{}

\begin{document}
	
	\thispagestyle{specialfooter}
	
	\noindent
	\textbf{\LARGE Reduced-order modeling of left ventricular flow subject to\\\\ aortic valve regurgitation}\hfill\break
	
	\begin{changemargin}{0.25in}{0.00in}
		{\large Di Labbio G$^{\dag}$, Kadem L}\hfill\break
		\textit{Laboratory of Cardiovascular Fluid Dynamics, Concordia University, Montr\'{e}al, QC, Canada, H3G 1M8}\hfill\break
		
		The present focus of heart flow studies is largely based on flow within the left ventricle and how this flow changes when subject to disease. However, despite recent advancements, a simple tractable model of even healthy left ventricular flow has not been produced and made available. Reduced-order modeling techniques, such as proper orthogonal decomposition (POD) and dynamic mode decomposition (DMD), offer an effective means of expressing the large datasets obtained from experiments or numerical simulations using low-dimensional models. While POD and DMD are often used to identify coherent structures in fluid dynamics, their use as a modeling tool has not found much merit in the cardiovascular flow community. In this work, we use POD and DMD to construct reduced-order models for a healthy left ventricular flow as well as for that under the influence of a particular disease shown to exhibit rich and unique intraventricular fluid dynamics, namely, aortic regurgitation (a leaking aortic valve). The performance of the two methods in reconstructing the intraventricular flows and derived quantities is evaluated, and the selected reduced-order models are made available.
		
		\hfill\break
		* \textit{Data pertaining to this article will be made available from the authors upon reasonable request.}\hfill\break
		* \textit{The authors have no conflicts of interest to declare.}\hfill\break
		* \textit{Please cite as}: Di Labbio, G., \& Kadem, L. (2019). Reduced-order modeling of left ventricular flow subject to aortic valve regurgitation. \textit{Physics of Fluids}, \textit{31}(3), 031901.
		
		\hfill\break
		\raisebox{0.95pt}{\small\textcopyright} 2019 Giuseppe Di Labbio \& Lyes Kadem. This manuscript version is made available under the AIP License to Publish Agreement with Authors, more information regarding usage terms can be found at \href{https://publishing.aip.org/resources/researchers/rights-and-permissions/permissions/}{\url{https://publishing.aip.org/resources/researchers/rights-and-permissions/permissions/}}. This article may be downloaded for personal use only. Any other use requires prior permission of the author and AIP Publishing. This article appeared in ``Di Labbio, G., \& Kadem, L. (2019). Reduced-order modeling of left ventricular flow subject to aortic valve regurgitation. \textit{Physics of Fluids}, \textit{31}(3), 031901.'' and may be found at \href{https://doi.org/10.1063/1.5083054}{\url{https://doi.org/10.1063/1.5083054}}. The supplementary information associated with this manuscript can be found along with the published article at the same link while the associated data is made available at \href{https://github.com/dilabbiog/ROMs--LV_flow_with_AR}{\url{https://github.com/dilabbiog/ROMs--LV_flow_with_AR}}.
	\end{changemargin}
	
	\section{\label{sec:Intro}Introduction}
		
		The human heart is a rather fascinating and surprisingly efficient four-chamber pump operating in marvelous synchrony. Nature has fashioned a pump that appears to be optimal in many respects and, to some extent, even adaptable to its own diseases. For a fluid dynamicist, the general perspective is that the flow in each chamber must be favorable for natural heart function and, therefore, functional changes brought on by a disease ought to be readily observed in the flow. The most instinctive notion by far is that the flows in healthy heart chambers minimize energy dissipation when compared to their diseased counterparts, requiring minimal work input on the part of the heart muscle to pump blood. Such was the hypothesis of \citeA{Kilner00}, for example, who discussed the natural swirling flows occurring in all healthy heart chambers with elegant magnetic resonance flow visualizations. Some evidence for this presumption has been provided over the past twenty years or so mainly in the context of the left ventricle, the heart's most laborious chamber; see the work of \citeA{Pedrizzetti05}, for instance. Indeed, the focus of heart flow studies has been largely on the left ventricle and how the corresponding intraventricular flow changes under various conditions such as with prosthetic valves or pathologies; the reader is referred to the work of \citeA{PedrizzettiDomenichini15} for a thorough review. Following this recent interest, this work focuses on flow in the left ventricle subject to a particular disease exhibiting rich and unique intraventricular fluid dynamics, namely, aortic valve regurgitation.
		
\begin{figure}[!t]
			\centering
			\includegraphics[scale=1]{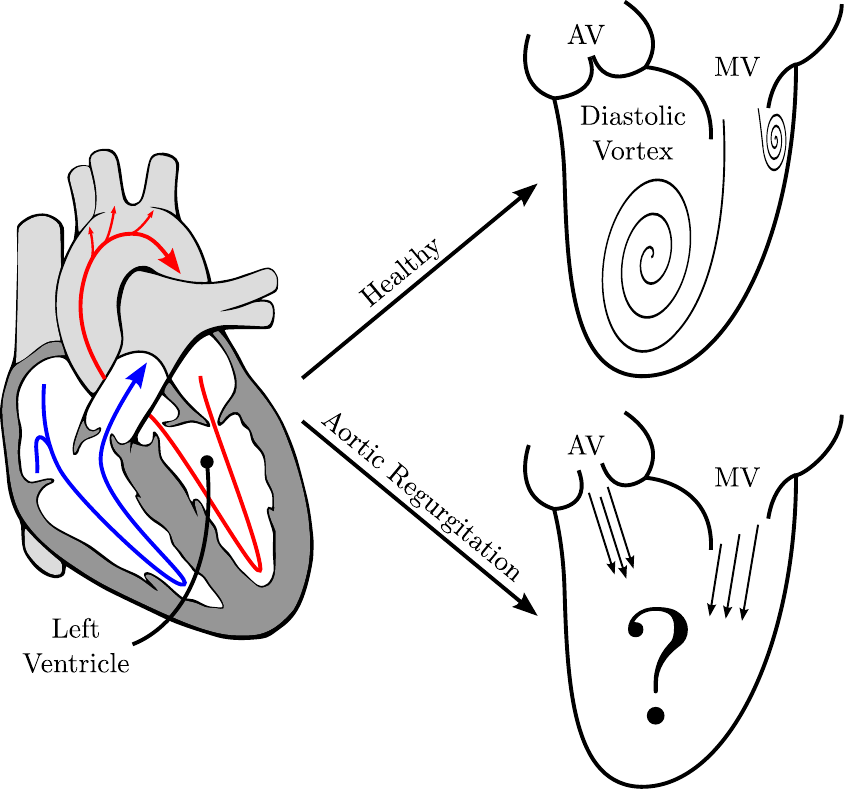}
			\caption{Schematic of the heart (left), with the left ventricle identified, showing the oxygenated (red) and deoxygenated (blue) blood flow paths. The filling flow in a healthy left ventricle compared to one with aortic valve regurgitation is also shown schematically. The image on the left was adapted from \href{https://commons.wikimedia.org/wiki/File:Heart_circulation_diagram.svg}{Patrick Lynch's illustration} under the \href{https://creativecommons.org/licenses/by-sa/2.5/}{CC BY-SA 2.5 License} (2006). Here and in the following figures, AV denotes aortic valve and MV mitral valve.}
			\label{fig:ProbSchem}
		\end{figure}

		Healthy left ventricular filling occurs strictly from the mitral valve, the eccentricity of which imparts a coherent swirl to the entire ventricular blood volume (see Fig.~\ref{fig:ProbSchem} for a schematic). By contrast, aortic regurgitation is characterized by a leaking aortic valve and so left ventricular filling occurs from two orifices, resulting in an interaction between two pulsatile jets in a confined, elastic geometry (see Fig.~\ref{fig:ProbSchem}). The interaction between two fluid jets has only been previously explored in the context of free or wall-bounded jets, mostly in the case of two parallel in-plane jets (cf.\ the work of \citeA{Bisoi17} and the references therein) and rather recently in the case of intersecting and somewhat free jets \cite{Houser18}. Aortic regurgitation in the left ventricle therefore poses a new and rather unique fluid mechanics problem. The flow has been previously investigated \textit{in vivo} by \citeA{Stugaard15}, who showed an increase in energy loss associated with increasing regurgitation severity. This has also been shown to occur \textit{in vitro} by \citeA{Okafor17} and \citeA{DiLabbioKadem18} including in earlier preliminary demonstrations by our group \cite{Raymondet16, DiLabbio16, BenAssa17}. Most recently, \citeA{DiLabbio17, DiLabbioVetelKadem18} investigated the flow from an enriching Lagrangian perspective, demonstrating various unique blood transport characteristics within the left ventricle using the finite-time Lyapunov exponent and maps of Lagrangian particle residence time. Nevertheless, whether the results of these studies can be observed \textit{in vivo} remains to be seen as there is currently no such database or literature with which to compare full left ventricular flow field data for aortic regurgitation of various grades. With this in mind, we believe it to be of utmost importance for researchers to have flow models available with which to compare future findings, whether arising from \textit{in silico}, \textit{in vitro}, \textit{in vivo}, or \textit{ex vivo} data. Furthermore, to date, no model intraventricular flow has been made available for even a healthy left ventricle. Here, we therefore look to construct and provide data-driven reduced-order models of the healthy and regurgitant intraventricular flows analyzed in our previous studies \cite{DiLabbioKadem18, DiLabbioVetelKadem18}. With this, both the fluid dynamics and clinical communities will have access to \textit{in vitro} data-driven flow models that capture the underlying physical phenomena and can be used for the purposes of comparison, reproduction, or further deduction.
		
		There are several data-driven reduced-order modeling techniques used in the fluid dynamics literature, the most popular being proper orthogonal decomposition (POD), which was first introduced by \citeA{Lumley67} and is now well-founded. Additionally, since its introduction by \citeA{Schmid08}, dynamic mode decomposition (DMD) has also had widespread use and success. Both techniques provide an effective means of reducing the large datasets acquired from experiments or numerical simulations to low-dimensional descriptions. Particularly, these two methods are promising for the reduced-order modeling of left ventricular flow fields. On the one hand, with these flows presumed to be well-described by their energetics in the literature, POD may be expected to produce optimal low-dimensional models. On the other hand, due to the periodic presence and propagation of distinct coherent structures in these flows (such as the diastolic vortex or regurgitant jet in Fig.~\ref{fig:ProbSchem}), DMD may be expected to describe these features with fewer modes given that the temporal dynamics of each mode comprises a single frequency of the governing flow. For a comprehensive review of DMD and other model reduction techniques (including POD), the reader is referred to the work of \citeA{Kutz16}, \citeA{Rowley17}, and \citeA{Taira17}. With regard to cardiovascular flows, the use of POD is limited to only a few studies and yet encompasses a range of purposes including the denoising of experimental data \cite{Cenedese05, Charonko10, Charonko13}, flow state estimation \cite{McGregor08, McGregor09, McLeod10, Guibert14, Manzoni12, Ballarin16}, and a general characterization of flow complexity \cite{Grinberg09, Kefayati13}. The application of DMD to such flows appears in far fewer studies but nonetheless has found use in the identification of coherent structures \cite{Delorme14}, in revealing bio-markers to characterize disease \cite{Borja16, Mikhail17} as well as in constructing reduced-order models \cite{Lozowy17}.
		
		With the focus of this study being to construct reduced-order models of the healthy and regurgitant intraventricular flows described in the work of \citeA{DiLabbioKadem18} and \citeA{DiLabbioVetelKadem18}, we begin by describing the experimental model from which the data were acquired in Sec.~\ref{sec:Methods}. For the unfamiliar reader, we define the snapshot POD \cite{Sirovich87a} and exact DMD \cite{Tu14} methods in Appendixes~\ref{app:PODAlg} and~\ref{app:DMDAlg}, respectively. In Sec.~\ref{sec:POD}, we apply POD to the datasets (Sec.~\hyperref[sec:PODconst]{\ref*{sec:POD}.\ref*{sec:PODconst}}) and evaluate the performance of the resulting reduced-order models to reconstruct them (Sec.~\hyperref[sec:PODperf]{\ref*{sec:POD}.\ref*{sec:PODperf}}). We continue in the same fashion for DMD throughout Sec.~\ref{sec:DMD}. We then offer some concluding remarks and recommendations in Sec.~\ref{sec:Conc}.
		
	\section{\label{sec:Methods}Experimental Model}
		
\begin{figure}[!b]
			\centering
			\includegraphics[scale=1]{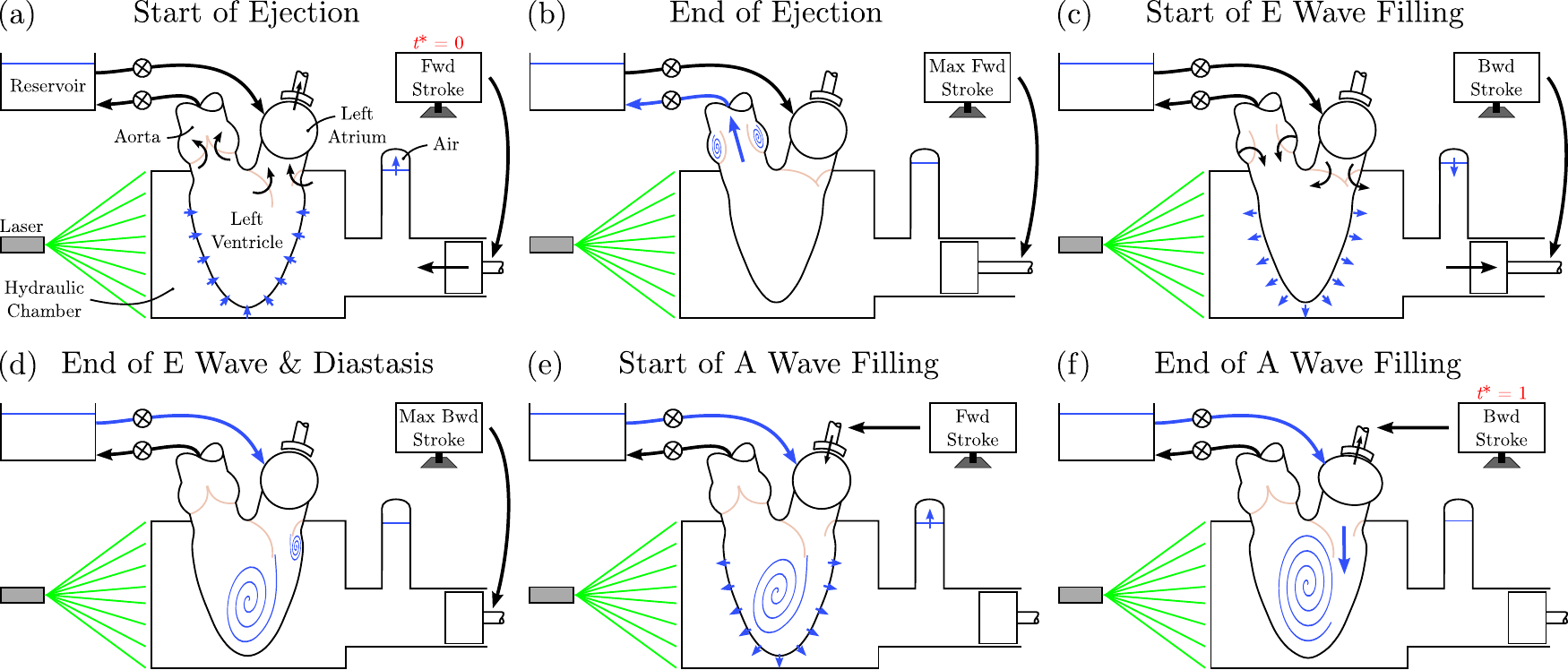}
			\caption{Operation of the double-activation left heart simulator over one complete cycle for the healthy scenario. (a)$\rightarrow$(b) The piston completes its forward stroke, compressing the left ventricle hydraulically for ejection of fluid up to the reservoir through the aortic valve. (c)$\rightarrow$(d) The piston completes its rearward stroke, expanding the left ventricle hydraulically to fill from the mitral valve for the \textit{E} wave and diastasis portions of the filling phase. Note that in the case of aortic regurgitation, filling would here also partly occur from the regurgitating aortic valve. (e)$\rightarrow$(f) The left atrium is compressed to provide the \textit{A} wave of filling, supplying an additional volume of fluid to the left ventricle while the piston remains still. The cycle then repeats.}
			\label{fig:Simulator}
		\end{figure}

		The methodology used to acquire the datasets being modeled in this work is largely described in the work of \citeA{DiLabbioKadem18} and \citeA{DiLabbioVetelKadem18}. Here, we provide an overview to place the applicability and limitations of the data into context and refer the reader to the aforementioned papers for further details. Aortic valve regurgitation was simulated on an in-house left heart simulator. We have previously shown the ability of this simulator to reproduce many healthy left heart flow phenomena in the work of \citeA{DiLabbioKadem18} and \citeA{DiLabbioVetelKadem18}, including aortic and left ventricular pressure waveforms as well as intraventricular flow behavior such as mitral vortex formation and propagation. The duplicator possesses a symmetric and optically clear silicone left ventricle, having a refractive index of $1.41 \pm 0.01$, which is held within an acrylic hydraulic chamber; the three-dimensional model of the ventricle has been made available as part of the supplementary material of \citeA{DiLabbioKadem18}. An anatomical model of the left atrium and a model aorta (complete with the sinuses of Valsalva) are also included in the flow circuit and made of the same silicone (SILASTIC\textsuperscript{TM} RTV-$4234$-T$4$, The Dow Chemical Company; USA), providing some compliance to the system. Compliant silicone rubber tubing, having $35$ Shore A hardness, is used for the remaining connections, namely, to join the left atrium and aorta to the reservoir (see Fig.~\ref{fig:Simulator}). Trileaflet bioprosthetic valves were used both in the aortic and mitral positions, having nominal diameters of $25$ and $23$~mm, respectively. The working fluid was a mixture of $60$\% water and $40$\% glycerol by volume at an operating temperature of $23.1 \pm 0.2$~$^{\circ}$C, having a refractive index of $1.39$ and a measured density ($1100$~kg/m$^3$) and dynamic viscosity ($4.2$~cP) not far from those of blood. The ventricle is activated hydraulically using a piston-cylinder arrangement, with the forward motion of the piston compressing the ventricle for ejection and rearward motion expanding the ventricle for the \textit{E} wave of the filling phase (the \textit{E} wave corresponds to filling due to the relaxation of the heart muscle). The piston is driven by an electromagnetic linear motor (LinMot, NTI AG; Switzerland) using a drive offering $32$-bit positional resolution (Servo Drive E$1100$-RS, NTI AG; Switzerland), which translates to a volumetric resolution of $0.2$~$\upmu$L for the system. The \textit{A} wave of filling, corresponding to ejection of blood from the left atrium into the left ventricle by atrial contraction, is decoupled from the \textit{E} wave by physically compressing the left atrium using a servomotor (Dynamixel RX-$24$F, Robotis; USA) acting on a simple cam-follower mechanism, having a maximum speed of $126$~rpm under no load and a positional resolution of $0.29^{\circ}$. The operation of the system over one complete cycle is detailed in Fig.~\ref{fig:Simulator} and its corresponding caption. This decoupling of the \textit{E} and \textit{A} waves of filling, known as double-activation, is rather important for the simulation of regurgitation from the aortic valve since it should occur forcefully during the \textit{E} wave while only passively during the \textit{A} wave. By comparison, having the piston-cylinder arrangement control both the \textit{E} and \textit{A} waves of filling, as is often the case in left heart simulators, would result in excess and unrealistic regurgitation during the \textit{A} wave.
		
		Aortic regurgitation was produced in the system by pulling apart the three leaflets of the bioprosthetic aortic valve to produce a central regurgitant orifice of known area; the reader is referred to the work of \citeA{DiLabbioKadem18} and the corresponding supplementary material for further details regarding the mechanism. Based on the clinical guidelines in the work of \citeA{Nishimura14}, five severities of aortic regurgitation, as well as one healthy scenario, were investigated. Normalized by the fully open geometric aortic valve area ($3.00$~cm$^2$), the simulated regurgitant orifice areas correspond to $0$\%, $3.3$\%, $5.9$\%, $8.5$\%, $17.2$\%, and $26.1$\%. As it is observed in patients with chronic aortic regurgitation \cite{Bekeredjian05}, the forward stroke volume was held constant at $64 \pm 4$~mL, corresponding to a cardiac output of $\sim4.5$~L/min. The peak aortic pressure was also held constant at $121 \pm 5$~mmHg along with the heart rate at $70$~beats per minute ($1.17$~Hz), corresponding to a cycle period of $T = 0.857$~s. Using the nominal mitral/aortic valve diameter $d$ as a length scale and incorporating the average flow rate $\dot{\volume}$ across the mitral/aortic valve into the velocity scale $U = \dot{\volume}/A = 4\dot{\volume}/{\pi}d^2$, we define the Reynolds and Womersley numbers as $\mathrm{Re} = 4{\rho}\dot{\volume}/({\pi\mu}d)$ and $\mathrm{Wo} = (d/2)\sqrt{2\pi\rho f/\mu}$, respectively, where $f$ denotes the frequency. The mean mitral inflow Reynolds number among all cases is therefore $1075 \pm 60$, and the associated Womersley number is $15.9$. Note that, for the given scales, we may also define a Strouhal number $\mathrm{St} = 2\mathrm{Wo}^2/{\pi}\mathrm{Re}$; however, given that the flows under study are pulsatile, we simply use the Reynolds and Womersley numbers to characterize them. A summary of the experimental flow conditions is provided in Tab.~\ref{tab:Simchars}. The listed Reynolds and Womersley numbers may find use in relating the results of the studied intraventricular flows to pulsatile or synthetic jets. The intraventricular flow fields were obtained using two-dimensional time-resolved particle image velocimetry (PIV) in the plane bisecting the mitral and aortic valve centers and the ventricle apex, namely, the plane shown in Fig.~\ref{fig:Simulator}. The flows were captured over one complete cardiac cycle. The simulator was left to run for $50$ cardiac cycles prior to each recording to ensure that any transient phenomena associated with system start-up were no longer present. The double-frame images were acquired $700$~$\upmu$s apart at an acquisition rate of $400$~Hz, the processing of which resulted in $343$ velocity field snapshots each having a final spatial resolution of $0.52$~mm $\times$ $0.52$~mm. Ten such time-resolved acquisitions were performed for each simulated case as a marker of the repeatability of the experiments. The uncertainty in the velocity fields captured by PIV was estimated to be around $5$\% of the maximum pointwise velocities for all simulated cases, ranging from $0.058$ to $0.086$~m/s. More information regarding the PIV acquisition and interrogation for this experiment can be found in the work of \citeA{DiLabbioVetelKadem18}. Throughout this and our previous studies, we make use of a nondimensionalized time $t^* = t/T$ defined such that $t^* = 0$ corresponds to the beginning of the ejection phase, $t^* = 0.438$ to the beginning of the filling phase, and $t^* = 1$ to the end of the filling phase (beginning of the ejection phase for the next cardiac cycle).
		
		\begin{table}[!h]
			\begin{center}
				\begin{tabular}{llcll}
					\multicolumn{5}{p{0.87\textwidth}}{Table \ref*{tab:Simchars}: Summary of experimental conditions.\hyperlink{note:Simchars}{\textsuperscript{a}} Reprinted with permission from G.\ Di Labbio, J.\ V\'{e}tel, and L.\ Kadem, \href{https://doi.org/10.1103/PhysRevFluids.3.113101}{Phys.\ Rev.\ Fluids} 3, 113101 (2018). Copyright 2018 American Physical Society.} \\
					\hline\hline\\[-1.0em]
					\multicolumn{2}{c}{Working fluid} & & \multicolumn{2}{c}{Simulator} \\
					\hline\\[-1.0em]
					Water-glycerol ratio & $60$-$40$ (by volume) & & Cardiac output & $4.5 \pm 0.3$~L/min \\
					Density ($\rho$) &  $1100$~kg/m$^3$ & & Cycle period ($T$) & $0.857$~s ($70$~bpm) \\
					Dynamic viscosity ($\mu$) & $0.0042$~Pa~s ($4.2$~cP) & & Forward stroke volume & $64 \pm 4$~mL \\
					Refractive index & $1.39$ & & Mitral inflow mean $\mathrm{Re}$ & $1075 \pm 60$ \\
					Temperature & $23.1 \pm 0.2$~$^{\circ}$C & & Mitral inflow peak $\mathrm{Re}$ & $3500 \pm 500$\\
					& & & Mitral inflow $\mathrm{Wo}$ & $15.9$ \\
					& & & Nominal mitral valve diameter & $23$~mm \\
					& & & Nominal aortic valve diameter & $25$~mm \\
					& & & Peak aortic pressure & $121 \pm 5$~mmHg \\
					\hline\\[-1.0em]
					\multicolumn{5}{c}{Regurgitation parameters} \\
					\hline\\[-1.0em]
					\multicolumn{2}{l}{Regurgitant orifice areas} & & \multicolumn{2}{l}{$0$\%, $3.3$\%, $5.9$\%, $8.5$\%, $17.2$\%, $26.1$\% (of $3.00$~cm$^2$)} \\
					\multicolumn{2}{l}{Diastolic aortic pressures} & & \multicolumn{2}{l}{$64$, $53$, $50$, $18$, $22$, $6$~mmHg} \\
					\multicolumn{2}{l}{Regurgitant fractions} & & \multicolumn{2}{l}{$0$, $0.11$, $0.34$, $0.40$, $0.47$, $0.52$} \\
					\multicolumn{2}{l}{Regurgitant inflow peak $\mathrm{Re}$} & & \multicolumn{2}{l}{$0$, $9970$, $14\, 400$, $11\, 300$, $12\, 100$, $9460$} \\
					\multicolumn{2}{l}{Regurgitant inflow $\mathrm{Wo}$} & & \multicolumn{2}{l}{$0$, $2.1$, $2.8$, $3.4$, $4.8$, $5.9$} \\
					\hline\hline\\[-1.9em]
					\multicolumn{5}{p{0.87\textwidth}}{\hypertarget{note:Simchars}{\textsuperscript{a}}For the regurgitant inflow, the orifice diameter was taken as $d = \sqrt{4A/\pi}$ for the calculation of the Reynolds ($\mathrm{Re}$) and Womersley ($\mathrm{Wo}$) numbers with $A$ being the regurgitant orifice area. For the mitral inflow, the nominal valve diameter was used. The regurgitant fraction is the regurgitant volume divided by the stroke volume.}
				\end{tabular}
			\end{center}
			\refstepcounter{table}\label{tab:Simchars}
		\end{table}
		
		In what follows, we apply POD and DMD to the velocity field data of each simulated case of aortic regurgitation. The goal is to offer a set of individualized reduced-order models tailored to the reconstruction of each specific case. Furthermore, in order to construct the models, we use the ensemble-averaged velocity fields of the ten time-resolved acquisitions made for each case; additional information regarding this decision is provided in Appendix~\ref{app:EnsAvg}. The models for each case make use of $343$ snapshots acquired at $400$~Hz, which we found to produce converged modes and temporal dynamics for both POD and DMD with the exception of the lowest energy modes in the case of POD and the highest frequency modes in the case of DMD. Similar numbers of snapshots and sample frequencies have also been found to give satisfactory results in other flows having similar velocity scales \cite{Kefayati13, Delorme14}. The MATLAB codes implementing the methods as well as the reconstruction algorithms are provided in the \href{ftp://ftp.aip.org/epaps/phys_fluids/E-PHFLE6-31-006903}{supplementary material}. Prior to proceeding with the data-driven modeling of the healthy and regurgitant intraventricular flows, it should be noted that the use of data reduction techniques for reconstructing intraventricular flows in general raises the rather interesting question of how to deal with moving or flexible boundaries. Here, we have simply used a constant rectangular flow domain among all snapshots of sufficient extent to contain the geometry throughout the entire cardiac cycle. For any given snapshot, the velocity of points falling outside the instantaneous ventricle boundary is identically zero. Evidently, using such an approach will result in modes that contain velocity vectors at points within the union of the left ventricular flow domains from all snapshots. Consequently, the reconstructed flow fields will contain false velocity vectors in all snapshots at points which lie outside their respective instantaneous ventricle boundary. Although we have found these false velocity vectors to be rather small in magnitude, we have nonetheless removed them in the flow reconstructions, by applying an instantaneous mask, to respect the motion of the ventricle walls; the reader is referred to the \href{ftp://ftp.aip.org/epaps/phys_fluids/E-PHFLE6-31-006903}{supplementary material} for a demonstration of how this is performed with the provided data.
		
	\section{\label{sec:POD}Modeling with Proper Orthogonal Decomposition}
		
		Here, we construct four POD models for each simulated intraventricular flow; we again refer the unfamiliar reader to Appendix~\ref{app:PODAlg} for the mathematical description of the method. The models are constructed according to the leading number of modes required to capture $98.0$\%, $99.0$\%, $99.5$\%, and $99.9$\% of the ensemble flow kinetic energy, the modes of course being ranked according to their kinetic energy content. For the purposes of flow reconstruction, we simply consider the projection of the modes onto the snapshot basis to evaluate the contribution of each mode at any given time as in Eq.~(\ref{eq:PODamps}), with the flow then being reconstructed from Eq.~(\ref{eq:PODlincomb}).
		
		\subsection{\label{sec:PODconst}Characteristics of POD applied to the intraventricular flows}
			
\begin{figure}[!h]
				\centering
				\subfloat[\label{fig:PODenergy}]{%
					\includegraphics{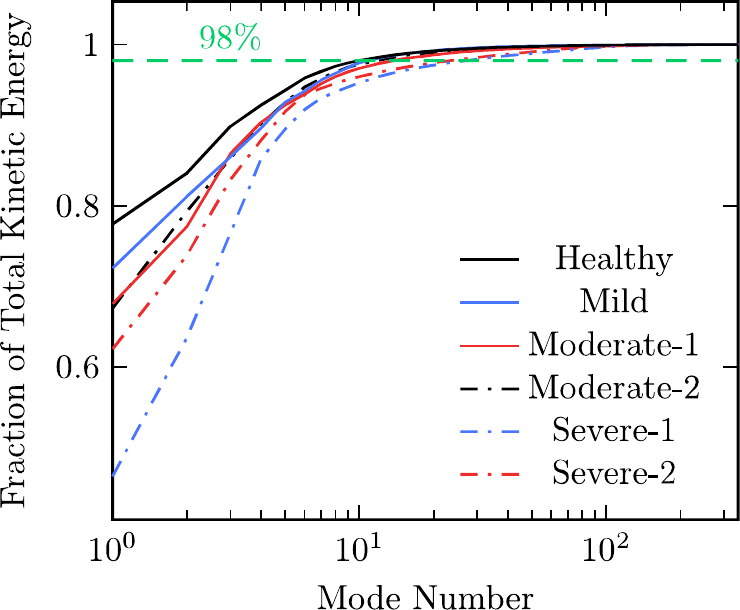}
				}\hfill
				\subfloat[\label{fig:PODentropy}]{%
					\includegraphics{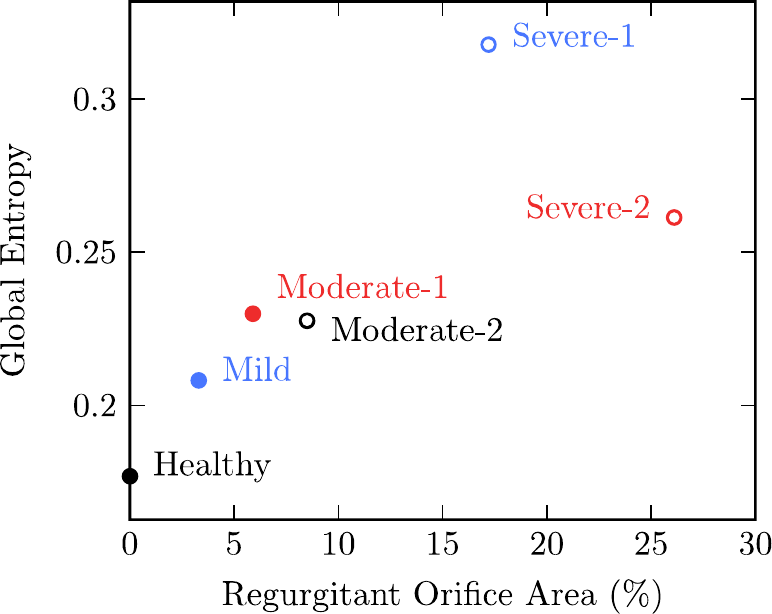}
				}
				\caption{(a) The fraction of accumulated kinetic energy against the mode number and (b) the Shannon entropy are shown for the proper orthogonal decompositions of each case.}
				\label{fig:PODchars}
			\end{figure}

			In this study, for regurgitant orifice areas of $0$\%, $3.3$\%, $5.9$\%, $8.5$\%, $17.2$\%, and $26.1$\% of the fully open geometric aortic valve area, the first proper orthogonal modes capture $77.7$\%, $72.3$\%, $67.8$\%, $67.3$\%, $46.4$\%, and $62.2$\% of their respective ensemble flow kinetic energy. In terms of energy content (i.e., the POD eigenvalues), the remaining dominant modes in all cases are often found to appear in pairs of relatively similar magnitude, which is often observed in the case of periodic flows due to the presence of traveling structures \cite{Kefayati13, Telib04, SantaCruz05, Bourguet09, Schlatter11}. For instance, for the healthy intraventricular flow, the second and third modes, respectively, make up $6.3$\% and $5.8$\% of the ensemble flow kinetic energy, while the fifth and sixth, respectively, make up $1.8$\% and $1.5$\%. The corresponding spatial modes do bare some resemblance to each other, although this becomes less apparent with increasing mode number. In terms of energy distribution, POD generally does give some crude indication of flow complexity in a global sense, suggesting that complex flows inherently require more modes to be adequately modeled up to a desired energy level. For instance, in order to capture $99.9$\% of the ensemble flow kinetic energy, the numbers of modes required in the reconstruction in order of increasing regurgitation severity are, respectively, $84$, $77$, $124$, $109$, $138$, and $125$; refer to Tab.~\ref{tab:PODchars} in Sec.~\hyperref[sec:PODperf]{\ref*{sec:POD}.\ref*{sec:PODperf}} for the number of modes required to capture $98.0$\%, $99.0$\%, and $99.5$\%. The corresponding accumulation of kinetic energy with the mode number is plotted in Fig.~\ref{fig:PODenergy} for all cases, showing a rapid convergence to $98.0$\% kinetic energy for all but the severe cases of aortic regurgitation ($\mathrm{ROA} = 17.2$ and $26.1$\%). Similar information is well-contained and condensed in the Shannon entropy of each decomposition in Fig.~\ref{fig:PODentropy}, defined as
			\begin{equation}
			\label{eq:PODentropy}
				H = -\frac{1}{\ln(n)}\sum_{j = 1}^n \frac{\lambda_j}{\mathrm{tr}(\boldsymbol{\Lambda})}\ln\left(\frac{\lambda_j}{\mathrm{tr}(\boldsymbol{\Lambda})}\right),
			\end{equation}
			with values closer to unity indicating a more disperse energy spectrum among the modes \cite{Aubry91}. There is a distinct increase in entropy between the healthy ($0.177$), mild ($0.208$), moderate ($0.230$, $0.228$), and severe ($0.318$, $0.261$) scenarios, suggesting more modes are generally required to reconstruct a given flow to within some pre-defined error with regurgitation severity category (i.e., the flows become increasingly complex with severity).
			
			\begin{figure}[!t]
				\centering
				\includegraphics{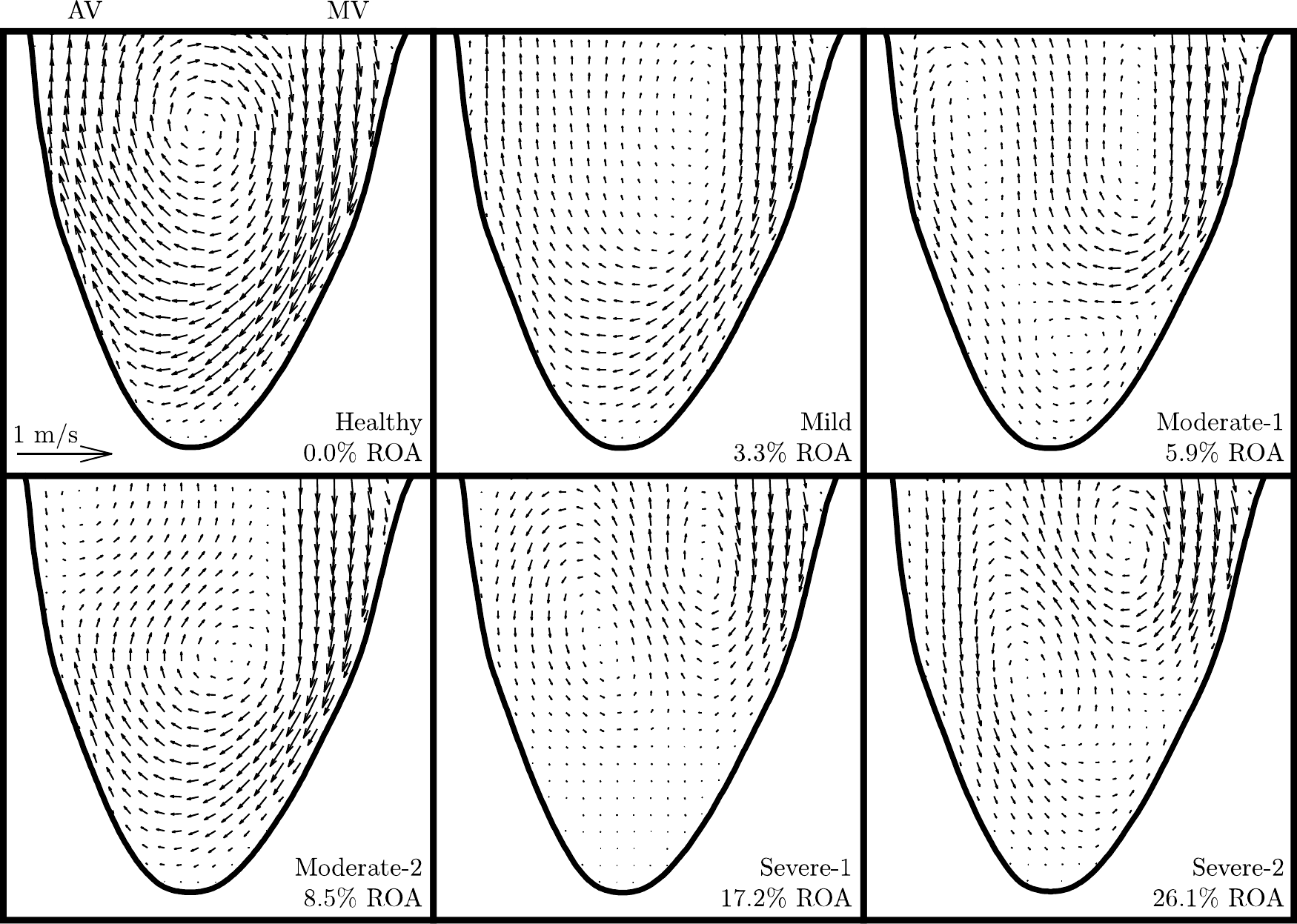}
				\caption{Time-averaged velocity field for all simulated cases of aortic regurgitation. The first proper orthogonal and dynamic modes effectively resemble the time-average for all cases.}
				\label{fig:TimeAvg}
			\end{figure}
			
			\begin{figure}[!t]
				\centering
				\includegraphics{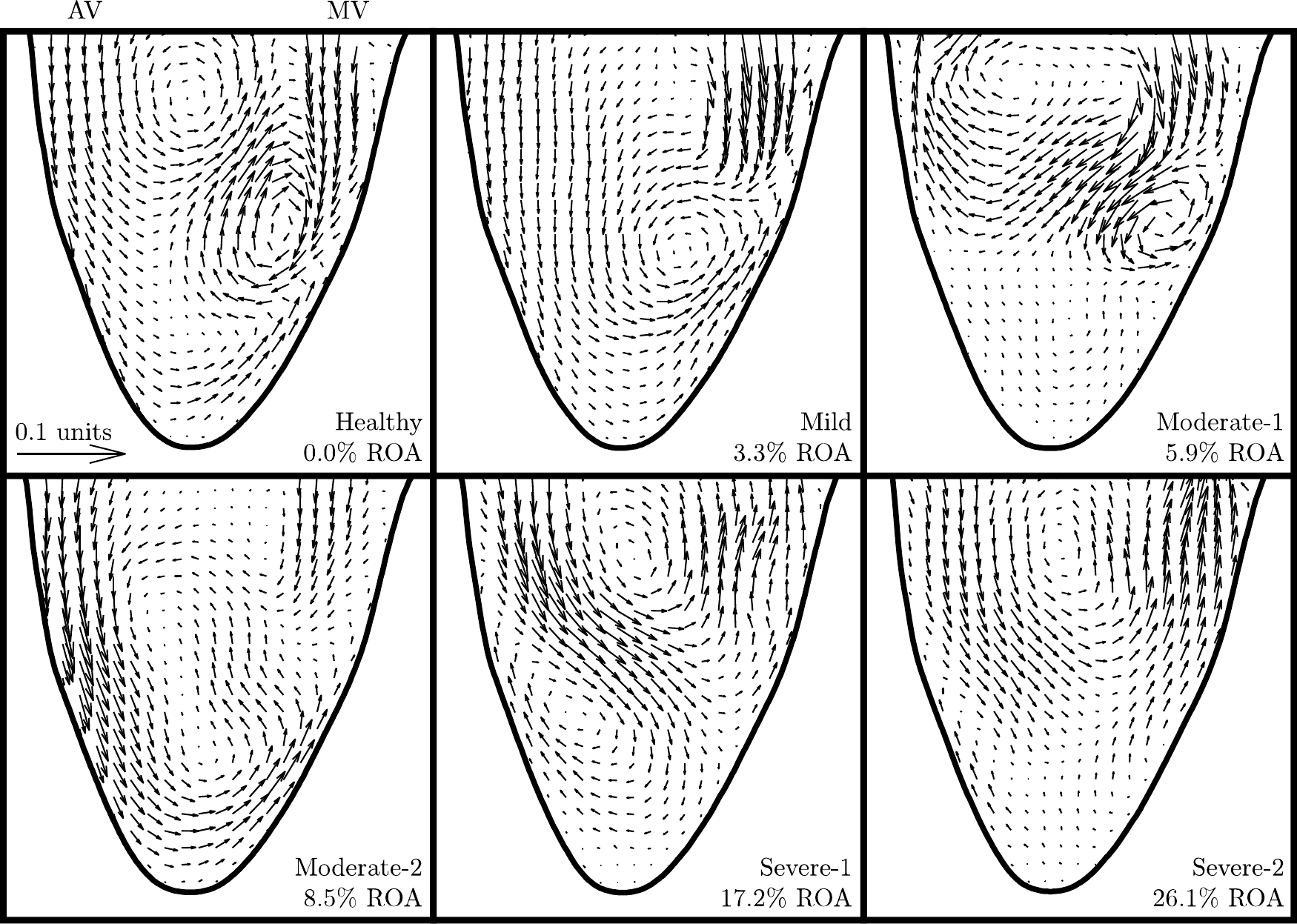}
				\caption{Second proper orthogonal modes for all simulated cases. These and several subsequent modes can be associated with the progression of the mitral vortex and regurgitant jet downstream.}
				\label{fig:PODM002}
			\end{figure}
			
\begin{figure}[!t]
				\centering
				\subfloat[\label{fig:PODamp1}]{%
					\includegraphics{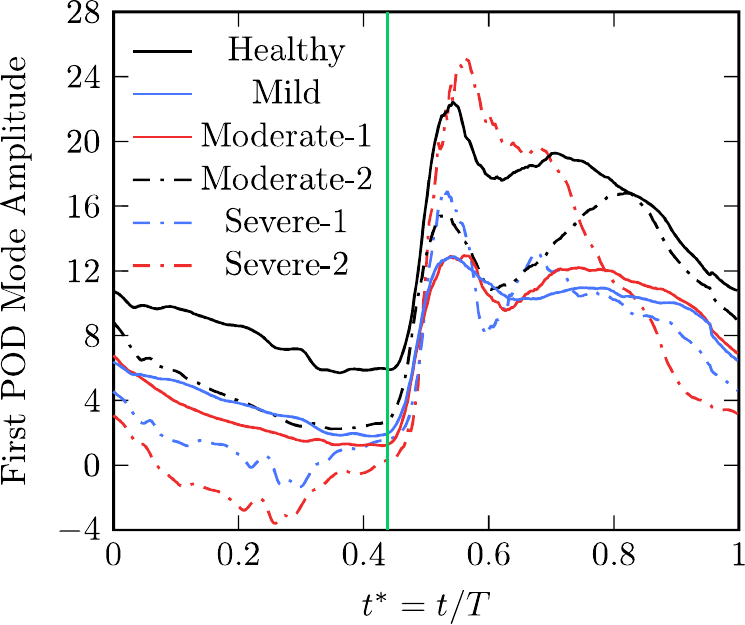}
				}\hfill
				\subfloat[\label{fig:PODamp2}]{%
					\includegraphics{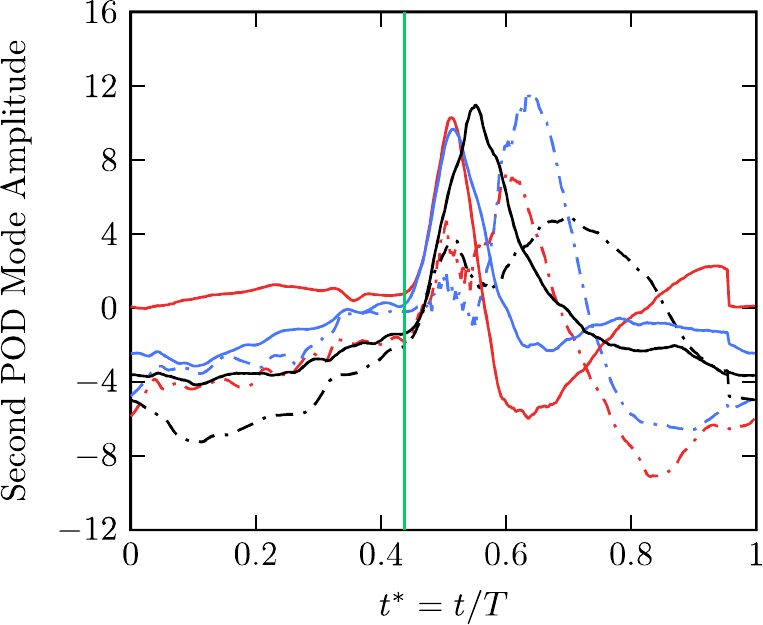}
				}
				\caption{Temporal dynamics of (a) the first modes and (b) the second modes from the proper orthogonal decompositions of each case. The vertical green line at $t^* = 0.438$ marks the beginning of the filling phase.}
				\label{fig:PODamps}
			\end{figure}

			Evidently, the decomposition places considerable emphasis on the filling phase in all cases, given that it is fluid dynamically the most energetically relevant part of the cycle. A significant portion of the modes can therefore be visually associated with filling dynamics. The time-averaged velocity fields, which are representative of the first proper orthogonal modes, are shown in Fig.~\ref{fig:TimeAvg} for all cases. These time-averages alone provide rather excellent descriptive pictures of the underlying filling flows discussed in the work of \citeA{DiLabbioKadem18} and \citeA{DiLabbioVetelKadem18}. For instance, the persistent clockwise swirl characteristic of a healthy left ventricle is clearly visible and occupies the entire ventricular domain (Fig.~\ref{fig:TimeAvg}, top left). With mild regurgitation (Fig.~\ref{fig:TimeAvg}, top center), this clockwise vortex was incapable of setting up in the ventricle's center, and so the time-average shows no distinct vortical pattern, capturing mainly the mitral inflow which is simply the most relevant dynamical feature in this particular case. The time-average also captures the distinct dynamics observed in the two moderate cases of aortic regurgitation. Namely, in the moderate-$1$ case (Fig.~\ref{fig:TimeAvg}, top right), the regurgitant jet manages to penetrate deeper into the ventricle while remaining close to the wall on the aortic side. By contrast, in the moderate-$2$ case (Fig.~\ref{fig:TimeAvg}, bottom left), the regurgitant jet sets up a counter-clockwise vortex opposing the upward ascent of the mitral inflow. Additionally, the time-average in the moderate-$1$ case reveals the small counter-clockwise vortex in the ventricle apex which has been shown to induce a laminar mixing region in the work of \citeA{DiLabbioVetelKadem18}, promoting an exceptionally high degree of blood stasis. For the two severe cases of regurgitation (Fig.~\ref{fig:TimeAvg}, bottom center and right), the dominant counter-clockwise vortex generated by the regurgitant jet is evident in both time-averages, and the clockwise mitral vortex is seen to be confined close to the mitral valve. Given the overall flow description available in Fig.~\ref{fig:TimeAvg}, the time-averaged flows over one complete cycle may be sufficient to judge the underlying filling dynamics in practice and therefore provides a useful bio-marker in the case of aortic regurgitation. The second modes, shown in Fig.~\ref{fig:PODM002}, display features associated with the progression of the mitral vortex and regurgitant jet downstream. The same could be said for the third and fourth modes, although not displayed here. The corresponding variation of the first and second mode amplitudes with time is shown in Figs.~\ref{fig:PODamp1} and~\ref{fig:PODamp2}, respectively. The influence of the first modes with time varies in a manner rather similar to the flow rate across the mitral valve, exhibiting little activity during ejection and larger amplitudes during filling. The amplitudes of the second modes, and the majority of the subsequent modes in fact, also exhibit little variation during the ejection phase and more activity during the filling phase.
			
			While it is often simply assumed that flow reconstruction using a POD capturing $98.0$\% or more of the ensemble flow kinetic energy provides a sufficient description of the flow, very few studies report the performance of the resulting reduced-order model in capturing secondary or derived flow phenomena, which is certainly of primary importance in modern fluid dynamics research. In general, constructing a data-driven reduced-order model of a flow has as an inherent consequence the effect of smoothing out the reconstructed flow field. While this effect is precisely what is desired in filtering applications, it may result in a loss of secondary information pertinent to the underlying dynamics if an insufficient number of modes are used in the reconstruction. As such, in what follows, we demonstrate the performance of the reduced-order models to reproduce the phenomena we have reported in the work of \citeA{DiLabbioKadem18} and \citeA{DiLabbioVetelKadem18}, such as the vorticity, energy, and material transport characteristics.
			
		\subsection{\label{sec:PODperf}Performance of the POD models}
			
			In the work of \citeA{DiLabbioKadem18}, it was shown that the intraventricular flow exhibits gradual vortex reversal during the filling phase with regurgitation severity and that the total energy loss per cycle due to viscous dissipation appears to increase linearly with regurgitant orifice area. As a measure of the total error for the former, Tab.~\ref{tab:PODchars} shows the integral of the circulation per unit area with respect to time during the filling phase ($T_f$), as defined by
			\begin{equation}
			\label{eq:Circ}
				\Gamma^* = \int_{T_f}\frac{\Gamma(t)}{A(t)}\mathrm{d}t = \int_{T_f}\left(\frac{1}{A(t)}\iint_A \left(\frac{\partial v}{\partial x} - \frac{\partial u}{\partial y}\right)\mathrm{d}A\right)\mathrm{d}t;
			\end{equation}
			note that this quantity is dimensionless. This circulation integral exhibits an overall increase from negative values (indicating clockwise rotation) to positive values (indicating counter-clockwise rotation) and therefore portrays the observed vortex reversal rather well. From Tab.~\ref{tab:PODchars}, it is clear that the overall vortical behavior of the flow is well-preserved by all four POD models. For the healthy intraventricular flow, the error decreases from $2.86$\% to $0.35$\% as more flow kinetic energy is preserved in the reconstruction. The largest errors are seen for the moderate-$1$ ($\mathrm{ROA} = 5.9$\%) and severe-$1$ ($\mathrm{ROA} = 17.2$\%) cases, decreasing from $15.6$\% to $1.34$\% for the former and $12.5$\% to $0.74$\% for the latter. While the viscous energy loss, defined by
			\begin{equation}
			\label{eq:VEL}
				\epsilon^* = \int_T\iint_A\epsilon(x, y, t)\mathrm{d}A\mathrm{d}t = \int_T\left(\frac{\mu}{2}\iint_A \sum_{\forall i,j}\left(\frac{\partial u_i}{\partial x_j} + \frac{\partial u_j}{\partial x_i}\right)^2\mathrm{d}A\right)\mathrm{d}t, \quad i,j \in \lbrace 1, 2 \rbrace,
			\end{equation}
			is not well-captured in magnitude by the ensemble-averaged flows (see Appendix~\ref{app:EnsAvg}), we show its corresponding reconstructed values in Tab.~\ref{tab:PODchars} nonetheless. The use of POD results in an underestimation, as expected, and it appears that the reconstruction to $99.9$\% kinetic energy is the only model that provides a satisfactory error, falling below $5$\% for all cases. However, the monotonically increasing trend in energy loss is preserved for the diseased cases for all attempted reconstructions, with the linearity being better represented starting with the POD model capturing $99.0$\% of the ensemble flow kinetic energy.
			
			\begin{table}[!b]
				\begin{center}
					\begin{tabular}{r*{15}{c}}
						\multicolumn{15}{l}{Table \ref*{tab:PODchars}: POD reconstruction characteristics and errors.\hyperlink{note:DMDchars}{\textsuperscript{a}}} \\
						\hline\hline\\[-1.0em]
						& \multicolumn{4}{c}{Number of Modes} & & \multicolumn{4}{c}{Circulation Error (\%)} & & \multicolumn{4}{c}{Energy Loss Error (\%)} \\
						\cline{2-5}
						\cline{7-10}
						\cline{12-15}\\[-1.0em]
						Captured KE (\%) $\rightarrow$ & $98.0$ & $99.0$ & $99.5$ & $99.9$ & & $98.0$ & $99.0$ & $99.5$ & $99.9$ & & $98.0$ & $99.0$ & $99.5$ & $99.9$ \\
						\hline\\[-1.0em]
						Healthy (\phantom{$0$}$0$\phantom{$.0$}\% $\mathrm{ROA}$) & $10$ & $17$ & $29$ & $84$  & & $2.86$ & $2.49$ & $1.12$ & $0.35$ & & $20.7$ & $16.6$ & $13.7$ & $4.63$ \\
						Mild (\phantom{$0$}$3.3$\% $\mathrm{ROA}$) & $11$ & $17$ & $26$ & $77$  & & $1.81$ & $1.52$ & $0.73$ & $0.32$ & & $20.7$ & $15.8$ & $11.4$ & $4.02$ \\
						Moderate-1 (\phantom{$0$}$5.9$\% $\mathrm{ROA}$) & $14$ & $24$ & $44$ & $124$ & & $15.6$ & $10.2$ & $5.07$ & $1.34$ & & $23.4$ & $17.8$ & $12.2$ & $2.97$ \\
						Moderate-2 (\phantom{$0$}$8.5$\% $\mathrm{ROA}$) & $12$ & $20$ & $36$ & $109$ & & $0.17$ & $1.78$ & $1.04$ & $0.24$ & & $23.7$ & $18.7$ & $14.0$ & $3.78$ \\
						Severe-1 ($17.2$\% $\mathrm{ROA}$) & $27$ & $54$ & $84$ & $138$ & & $12.5$ & $6.20$ & $0.04$ & $0.74$ & & $33.8$ & $21.5$ & $11.0$ & $2.24$ \\
						Severe-2 ($26.1$\% $\mathrm{ROA}$) & $23$ & $46$ & $71$ & $125$ & & $4.91$ & $2.44$ & $1.51$ & $0.26$ & & $36.1$ & $22.7$ & $11.5$ & $2.54$ \\
						\hline\hline\\[-1.0em]
						\multicolumn{13}{l}{\hypertarget{note:DMDchars}{\textsuperscript{a}}Here and in Tab.~\ref{tab:DMDchars}, KE denotes kinetic energy.}
					\end{tabular}
				\end{center}
				\refstepcounter{table}\label{tab:PODchars}
			\end{table}
			
\begin{figure}[!t]
				\centering
				\includegraphics{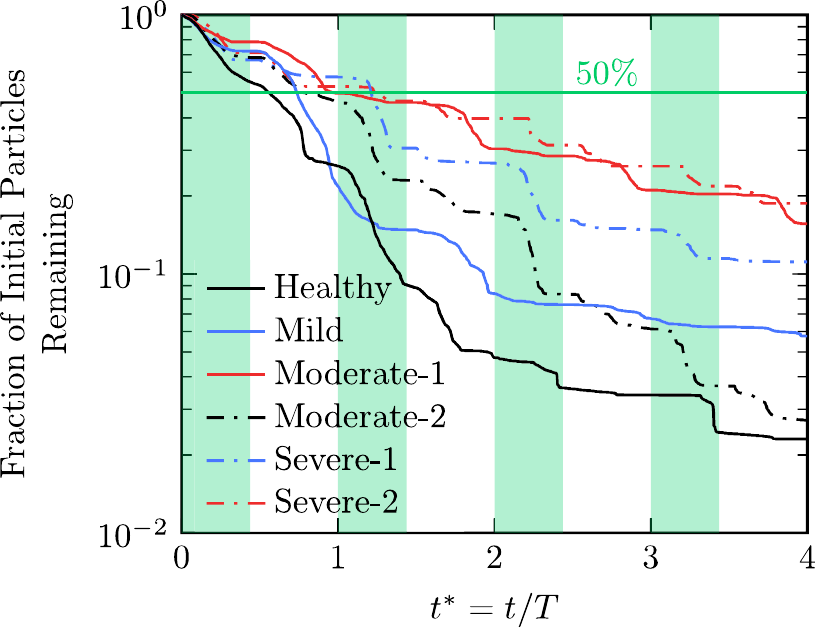}
				\caption{Approximately $800\, 000$ particles were released at the start of the ejection phase ($t^* = 0$) in the left ventricle and advected for four cardiac cycles. The plot shows the fraction of initial particles remaining with time. The regions shaded in green correspond to the ejection phases.}
				\label{fig:EnsAvgAdv}
			\end{figure}

			In the work of \citeA{DiLabbioVetelKadem18}, aortic regurgitation was investigated from a Lagrangian perspective and hence much of the results are dependent upon particle advection patterns. To evaluate the error involved, the flow was filled with $\sim800\, 000$ virtual particles at the start of the ejection phase and advected for four cycles. The advection was performed using the fourth-order Runge-Kutta scheme for time-stepping and bicubic interpolation to determine the velocities at the advected particle positions. As in the work of \citeA{DiLabbioVetelKadem18}, the same cycle was simply appended to itself as many times as needed for the advection. In order to evaluate error, the fraction of particles remaining in the ventricle with time is used. This provides some indication on blood stasis within the left ventricle, the behavior of which is shown in Fig.~\ref{fig:EnsAvgAdv} for the ensemble-averaged flows. By the end of the first ejection phase (at $t^* = 0.438$), the healthy left ventricle has ejected $45$\% of the initial particles, while only $27.5$\%, $21.5$\%, $31.6$\%, $33.0$\%, and $28.6$\% were ejected with aortic regurgitation in order of increasing severity. At an energy level of $98.0$\%, the POD models show an error in the fraction of remaining particles of $3.4$\% for the healthy scenario and of less than $1$\% for the regurgitant cases. At an energy level of $99.0$\%, the error reduces to $0.2$\% for the healthy scenario, and at $99.9$\%, the errors fall below $0.14$\% for all cases. Over two cardiac cycles, the errors propagate forward in time and increase considerably, surpassing $10$\% error for the healthy left ventricular flow reconstructed using $98.0$\% and $99.5$\% of its ensemble flow kinetic energy. Rather interestingly, the healthy left ventricular flow exhibits the most difficult global particle advection behavior to reconstruct. This is not due to the particle positions being in greater error than those of the regurgitant cases but rather to the very low degree of stasis exhibited by the healthy left ventricle, as shown in the work of \citeA{DiLabbioVetelKadem18}. Therefore, the small fraction of remaining particles promotes larger error values compared to the large fraction of particles remaining in the diseased cases, as elucidated in Fig.~\ref{fig:EnsAvgAdv}. Nonetheless, at an energy level of $99.9$\%, the POD models show errors falling below $2$\% for all cases over two cardiac cycles and below $4$\% over four cardiac cycles.
			
			With regard to the healthy intraventricular flow, the POD model capturing $99.0$\% of its ensemble flow kinetic energy performs rather well with the exception of underestimating the computed viscous energy loss and capturing global particle advection behavior for over one cardiac cycle. If advection behavior for over one cardiac cycle is of interest, we are inclined to suggest that only the POD model capturing $99.9$\% of the energy is sufficient. With regard to the regurgitant cases, the POD models capturing $98.0$\% of the energy appear to be sufficient in terms of advection behavior within one cardiac cycle whereas those capturing at least $99.5$\% of the energy would be suggested if additional cycles are of interest. We now move to constructing reduced-order models using DMD and compare their performance against those constructed using POD.
			
	\section{\label{sec:DMD}Modeling with Dynamic Mode Decomposition}
		
		The application of POD to velocity field data ranks modes based on their kinetic energy content, the underlying assumption being that the flow can be well-described by its energetics alone. By contrast, DMD places emphasis on the actual temporal dynamics, namely, how the flow field at one time instant is mapped directly from the flow history. Its modes represent single-frequency contributions to the original flow field, which may additionally grow or decay with time and have an associated phase. It will be interesting to see whether the intraventricular flows are better described in this context than in terms of kinetic energy.
		
		In order to have some means of comparison with the performance of the DMD models relative to those of POD, we again base the models on capturing $98.0$\%, $99.0$\%, $99.5$\%, and $99.9$\% of the ensemble flow kinetic energy. Recall that for DMD, the temporal dynamics $g_j(t)$ of mode $j$ is given by the $j$th row of the product $\mathbf{B}\mathbf{T}$ in Eq.~(\ref{eq:DMDapprox}) and therefore the energy of each mode is given by the integral of $g_j(t)$ over the cardiac cycle, namely,
		\begin{equation}
		\label{eq:DMDEnergy}
			E_j = \int_T |g_j(t)|^2 \mathrm{d}t.
		\end{equation}
		The ensemble flow kinetic energy is then given by the sum of all $E_j$. Here, we apply the exact DMD method of \citeA{Tu14} to the datasets; we again refer the unfamiliar reader to Appendix~\ref{app:DMDAlg} for the mathematical description of the method. The models are constructed according to the leading number of modes required to capture the desired energy level, with the modes first being sorted in terms of their amplitude and the flows being reconstructed from Eq.~(\ref{eq:DMDreconstruct}). Furthermore, strictly for the purposes of mode sorting, we found it necessary to premultiply the amplitudes by their respective Ritz values raised to the power of $n$ (the number of snapshots); i.e., the quantity $\left|\lambda_k^nb_k\right|$ was used for sorting, where $\lambda_k$ is given by Eq.~(\ref{eq:DMDeigvals}) and $b_k$ by Eq.~(\ref{eq:DMDamps}). This served to penalize modes having large amplitudes but weak contributions to the dynamics, as discussed in the work of \citeA{Tu14}.
		
		\subsection{\label{sec:DMDconst}Characteristics of DMD applied to the intraventricular flows}
			
\begin{figure}[!t]
				\centering
				\subfloat[\label{fig:DMDattract0}]{%
					\includegraphics{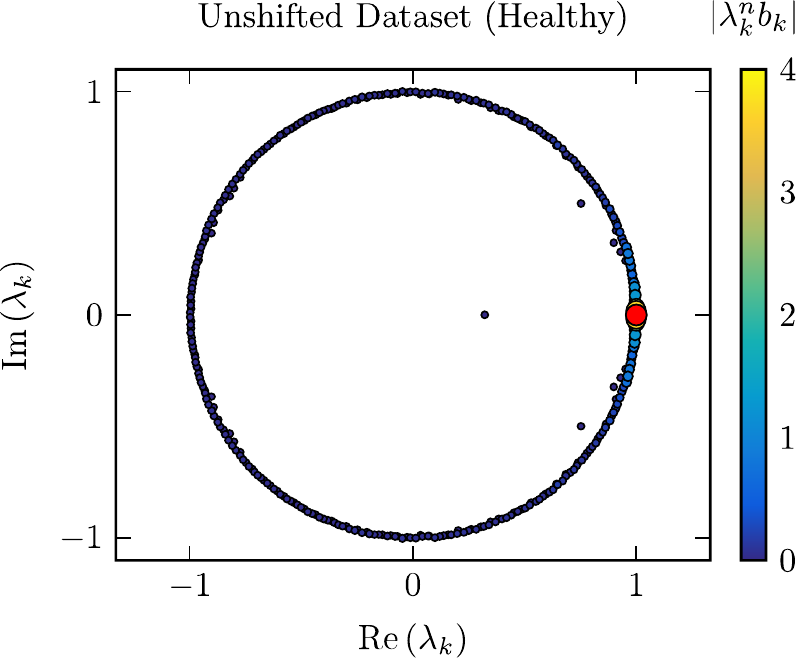}
				}\hfill
				\subfloat[\label{fig:DMDattract20}]{%
					\includegraphics{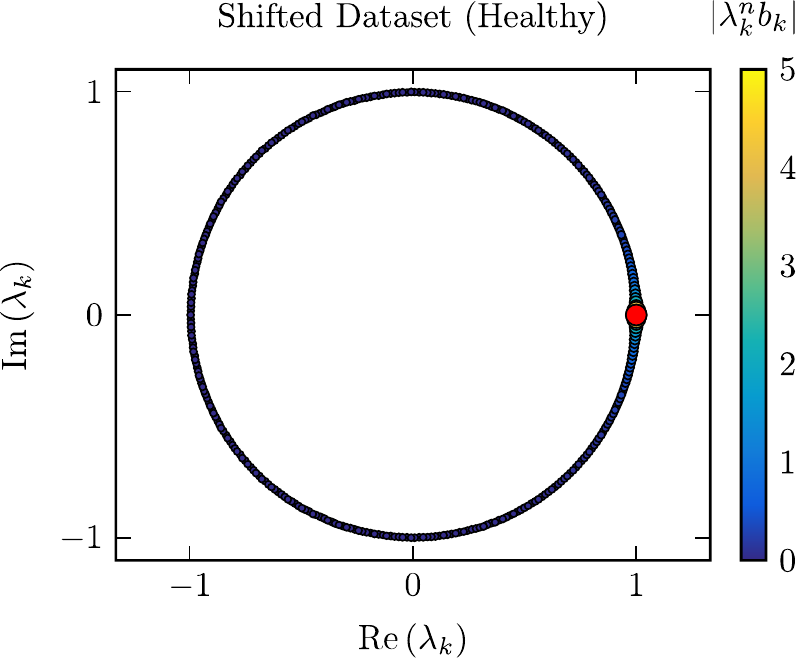}
				}\\
				\subfloat[\label{fig:DMDattract5}]{%
					\includegraphics{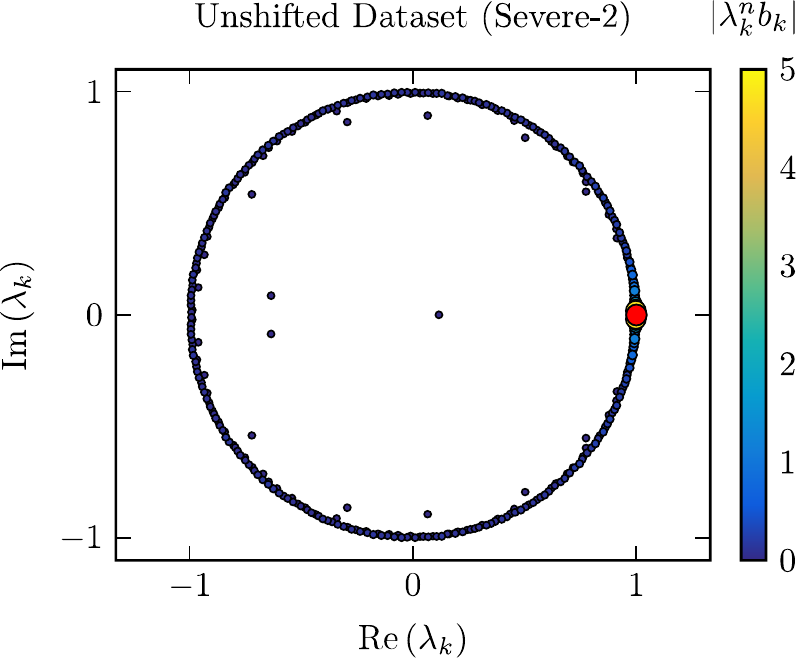}
				}\hfill
				\subfloat[\label{fig:DMDattract25}]{%
					\includegraphics{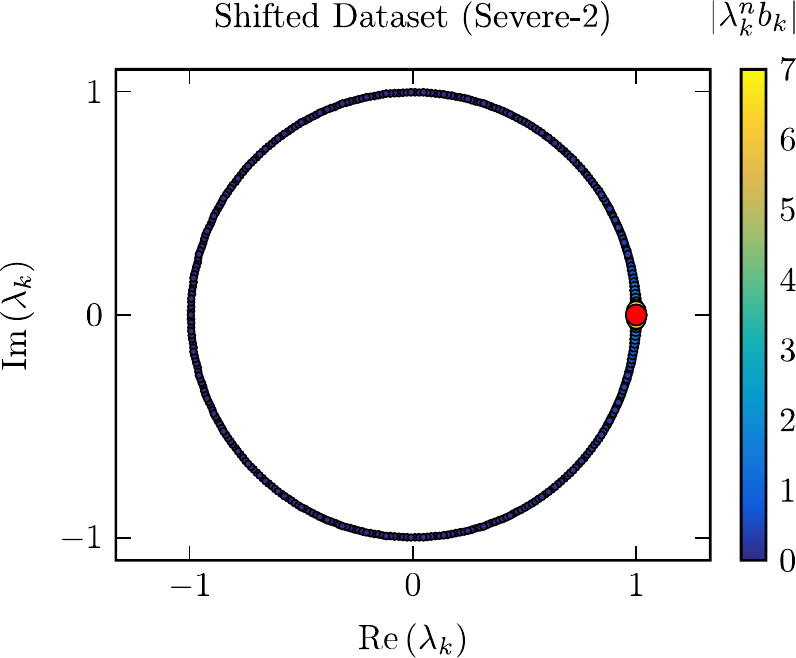}
				}
				\caption{Discrete-time eigenvalues corresponding to the DMD of the unshifted datasets for (a) the healthy and (c) most severe scenarios. The respective eigenvalues for the datasets shifted according to Eq.~(\ref{eq:ordering}) are shown in (b) and (d). The color and size of the points vary according to the penalized amplitudes ($|\lambda_k^nb_k|$) of the modes except for the first modes, which are shown in red and have amplitudes of $13.29$ in (a), $12.62$ in (b), $8.53$ in (c), and $7.36$ in (d).}
				\label{fig:DMDattract}
			\end{figure}

			In our first attempt to apply DMD to the datasets, we have noted the presence of dominant unstable modes appearing in the eigenvalue spectra for all cases. Such unstable modes are particularly undesirable for the purposes of modeling periodic flows as they will continue to grow without bound, causing the resulting models to fail for long times (i.e., in later cycles). We have also noted that while a significant number of modes were in fact stable, they decayed rather strongly in time. These strongly decaying modes are also unfavorable when modeling periodic flows as they will no longer influence the dynamics for sufficiently long times. Indeed, as discussed in the work of \citeA{Rowley09}, in the case of perfect periodicity, the Ritz values (discrete-time eigenvalues) ought to correspond to roots of unity and therefore the temporal dynamics ought to appear as pure sinusoids which neither grow nor decay in time; this is equivalent to saying that the Vandermonde matrix given by Eq.~(\ref{eq:Vand}) would be identical to the discrete Fourier transform matrix. Alas, the discordance between what was observed in the eigenvalue spectra and what is expected for periodic flows is in fact due to the experimental nature of the data. More precisely, it is due to the slight dissimilarity between the first and last snapshots of the datasets, which of course should represent a full cycle to within $0.3$\% error ($2$-$3$~ms) given the experimental conditions (cf.\ Sec.~\ref{sec:Methods}). However, by applying a shift to the ordering of the snapshot matrix, the first and last snapshots will be in better correspondence and therefore the periodicity of the data for the purposes of DMD will be improved. We have therefore shifted the datasets such that the first snapshot now corresponds to the start of the filling phase; i.e., we have performed the shift
			\begin{equation}
			\label{eq:ordering}
				\mathbf{X}_1^n =
				\begin{bmatrix}
					\textcolor{SpringGreen3}{\mid} & & \textcolor{RoyalBlue1}{\mid} & \textcolor{RoyalBlue1}{\mid} & & \textcolor{Firebrick2}{\mid} \\
					\textcolor{SpringGreen3}{\mathbf{x}_1} & \cdots & \textcolor{RoyalBlue1}{\mathbf{x}_{j-1}} & \textcolor{RoyalBlue1}{\mathbf{x}_j} &\cdots & \textcolor{Firebrick2}{\mathbf{x}_{n}} \\
					\textcolor{SpringGreen3}{\mid} & & \textcolor{RoyalBlue1}{\mid} & \textcolor{RoyalBlue1}{\mid} & & \textcolor{Firebrick2}{\mid}
				\end{bmatrix} \rightarrow
				\begin{bmatrix}
					\textcolor{RoyalBlue1}{\mid}         &        & \textcolor{Firebrick2}{\mid}         & 	\textcolor{SpringGreen3}{\mid}         &       & \textcolor{RoyalBlue1}{\mid}             \\
					\textcolor{RoyalBlue1}{\mathbf{x}_j} & \cdots & \textcolor{Firebrick2}{\mathbf{x}_n} & 	\textcolor{SpringGreen3}{\mathbf{x}_1} &\cdots & \textcolor{RoyalBlue1}{\mathbf{x}_{j-1}} \\
					\textcolor{RoyalBlue1}{\mid}         &        & \textcolor{Firebrick2}{\mid}         & 	\textcolor{SpringGreen3}{\mid}         &       & \textcolor{RoyalBlue1}{\mid}
				\end{bmatrix},
			\end{equation}
			where snapshot $j$ corresponds to the start of the filling phase. There is a subtle point of note here, namely, that the temporal spacing between all snapshots is $2.5$~ms, whereas that between snapshot $n$ and $1$ is only $2.1$~ms, which introduces some error in the DMD. Nonetheless, with this shift, the Ritz values move more closely toward the unit circle, which we show for the healthy and most severe cases in Fig.~\ref{fig:DMDattract}.
			
			\begin{figure}[!t]
				\centering
				\subfloat[\label{fig:DMDritz2AR0}]{%
					\includegraphics{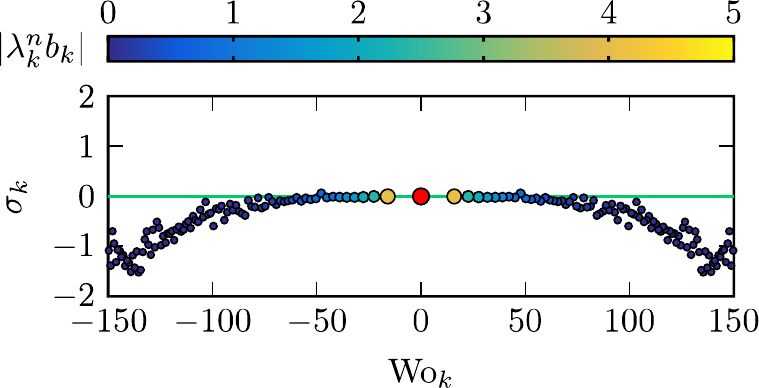}
				}\hfill
				\subfloat[\label{fig:DMDritz2AR1}]{%
					\includegraphics{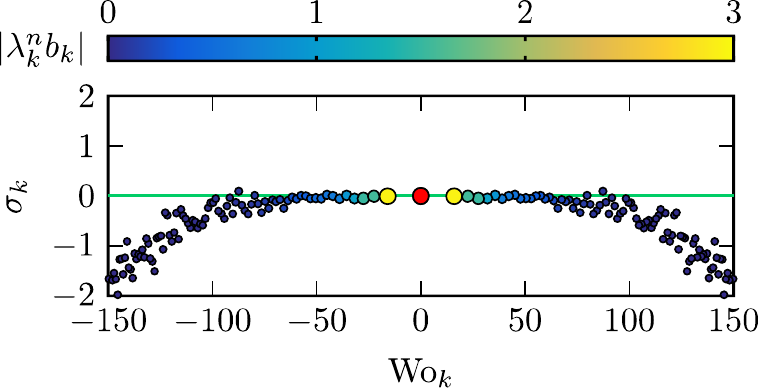}
				}\\
				\subfloat[\label{fig:DMDritz2AR2}]{%
					\includegraphics{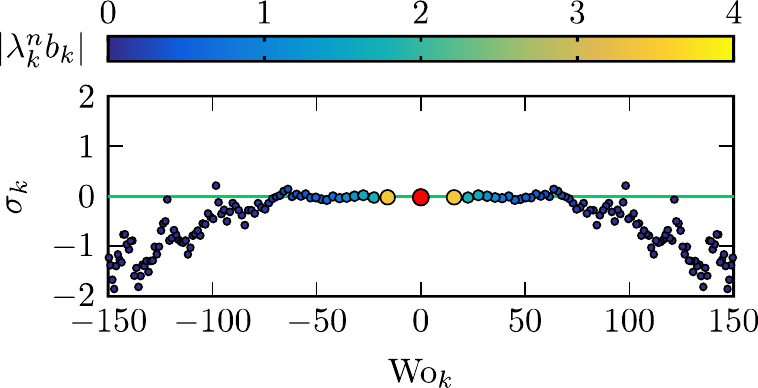}
				}\hfill
				\subfloat[\label{fig:DMDritz2AR3}]{%
					\includegraphics{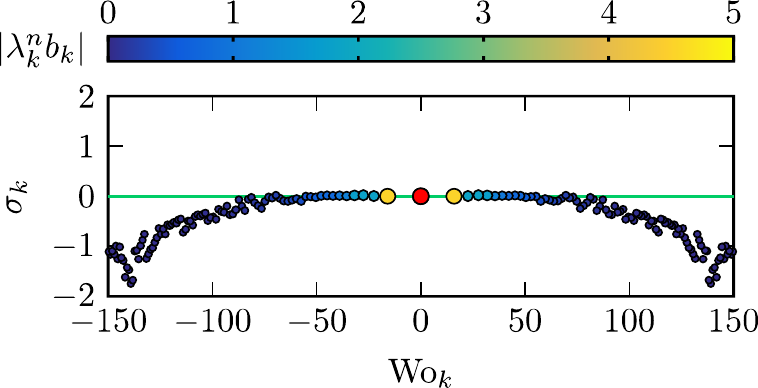}
				}\\
				\subfloat[\label{fig:DMDritz2AR4}]{%
					\includegraphics{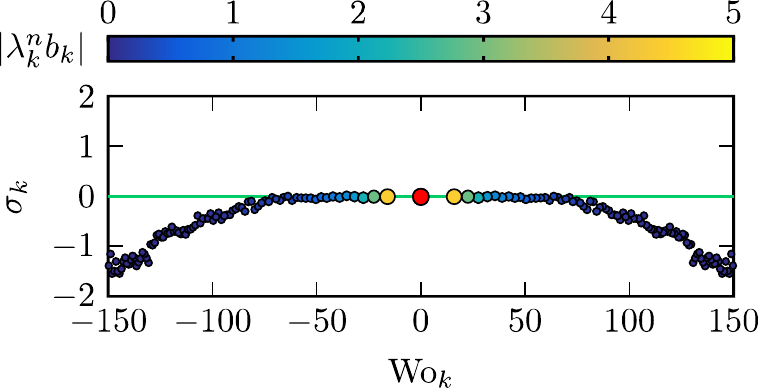}
				}\hfill
				\subfloat[\label{fig:DMDritz2AR5}]{%
					\includegraphics{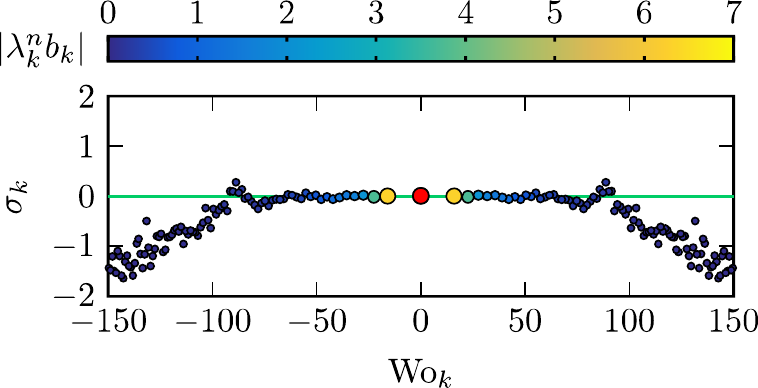}
				}
				\caption{Continuous-time eigenvalues ($\gamma_k = \ln\left(\lambda_k\right)/\Delta t$; see Appendix~\ref{app:DMDAlg}) for (a) healthy left ventricular flow as well as five severities of aortic regurgitation including (b) mild ($\mathrm{ROA} = 3.3$\%), (c,d) moderate ($\mathrm{ROA} = 5.9$ and $8.5$\%) and (e,f) severe ($\mathrm{ROA} = 17.2$ and $26.1$\%). The first modes, shown in red, have amplitudes of $12.62$, $7.17$, $6.92$, $9.12$, $6.08$, and $7.36$ in order of increasing regurgitation severity.}
				\label{fig:DMDritz}
			\end{figure}
			
			With this shift, the logarithmic mappings of the Ritz values (continuous-time eigenvalues) are now shown in Fig.~\ref{fig:DMDritz} for all cases in terms of the Womersley numbers of the modes; i.e., $\mathrm{Wo}_k = \mathrm{sign}(\omega_k)(d/2)\sqrt{\rho|\omega_k|/\mu}$ with $d$ selected as the nominal mitral valve diameter. The color and size of the continuous-time eigenvalues are scaled according to their penalized amplitudes. The points corresponding to the first modes are marked in red and are purposefully made not to follow the scale due to them generally having significantly larger amplitudes in comparison with the other modes. The penalized amplitudes of the first modes are given by $12.62$, $7.17$, $6.92$, $9.12$, $6.08$, and $7.36$ in order of increasing regurgitation severity. Note that in all cases, the most dominant modes have nearly zero real part, suggesting purely sinusoidal temporal dynamics without growth or decay. Only at higher frequencies, do we observe modes with growing or decaying temporal dynamics; however, many of these modes will nonetheless be ignored in the reconstructions due to their small penalized amplitude, improving the long-time behavior of the DMD models. Furthermore, given that the sampling rate of the datasets was $400$~Hz, frequencies above $200$~Hz ($|\mathrm{Wo}_k| > 208$), the Nyquist frequency, are not captured by the DMD but are nonetheless likely irrelevant to the dynamics in terms of penalized amplitude. In fact, the largest amplitudes are concentrated in the low frequency regime up to $\mathrm{Wo}_k \approx 66$ ($20$~Hz), about $4$ times the Womersley number of the mitral inflow, where they reduce to $2$\% to $5$\% of their maximum values. The magnitudes also fall off rather rapidly from the first mode, decreasing to between $1/3$ and $1/2$ of its value for the first mode pair for all but the severe cases ($\mathrm{ROA} = 17.2$ and $26.1$\%).
			
			\begin{figure}[!h]
				\centering
				\includegraphics{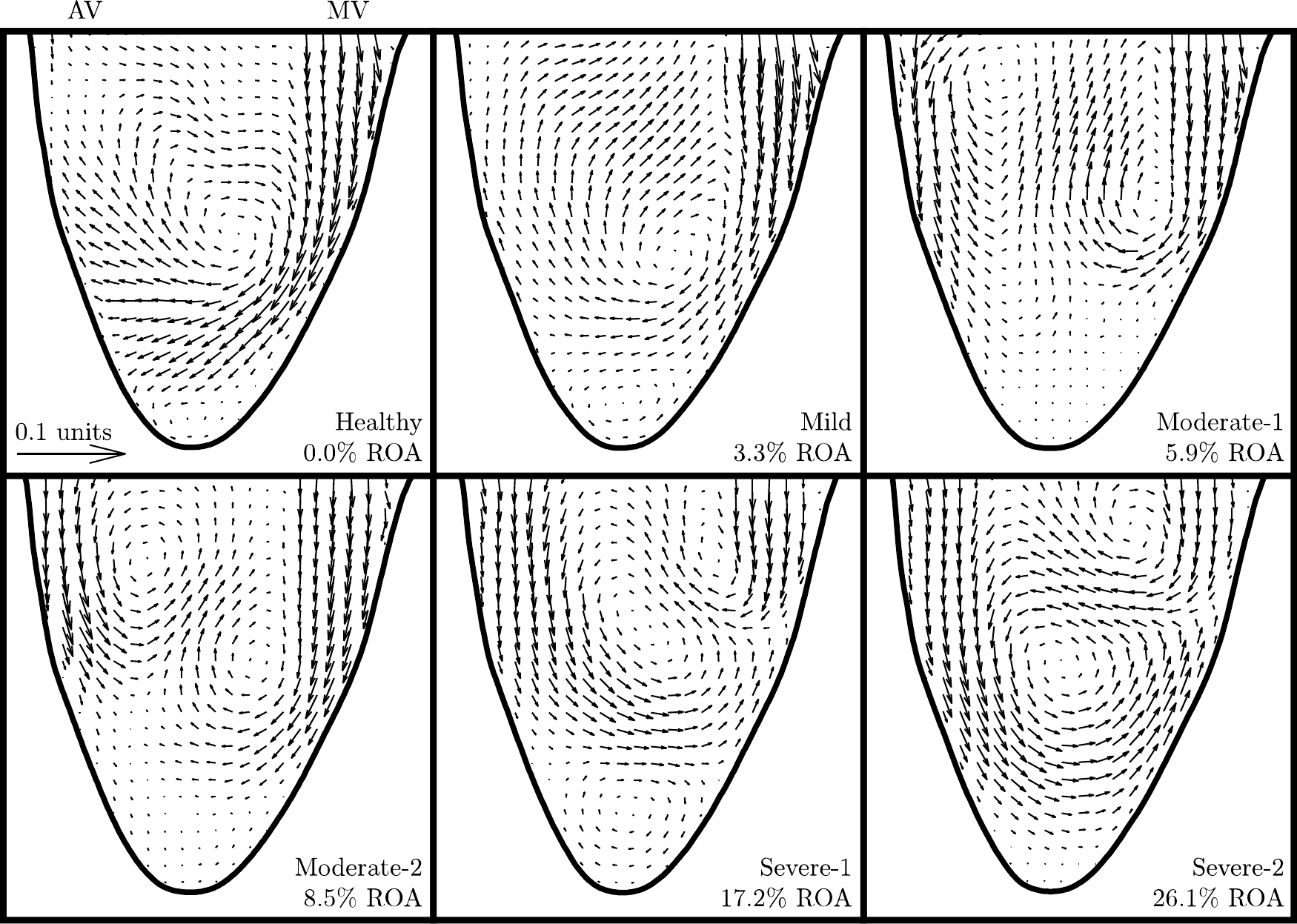}
				\caption{Real part of the first pair of dynamic modes (i.e., modes $2$ and $3$) for all simulated cases. The real part of the modes appears to be a direct progression of the mitral and regurgitant jets downstream along with their generated vortices.}
				\label{fig:DMDM002Re}
			\end{figure}
			
			\begin{figure}[!t]
				\centering
				\includegraphics{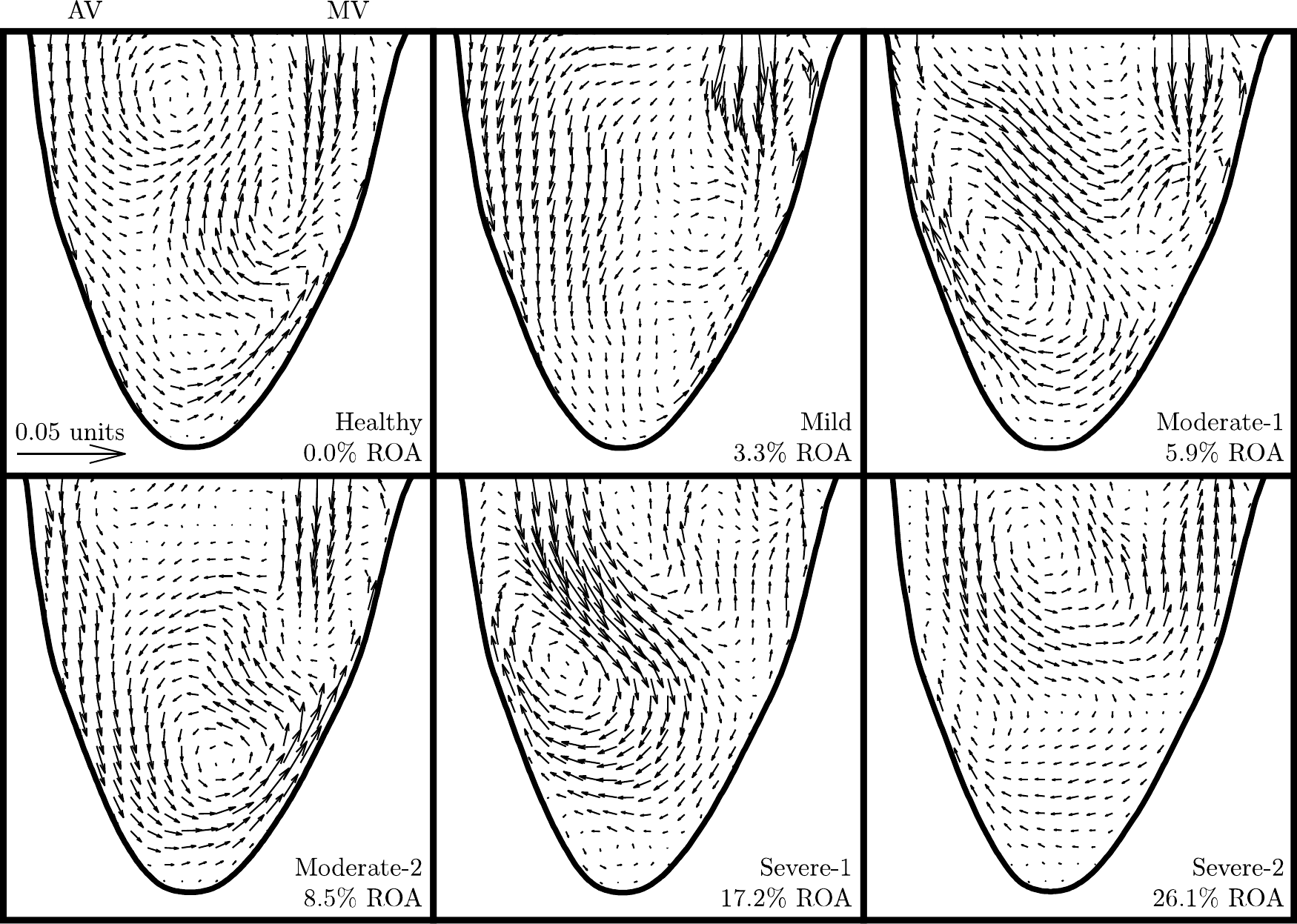}
				\caption{Imaginary part of the first pair of dynamic modes (i.e., modes $2$ and $3$) for all simulated cases.}
				\label{fig:DMDM002Im}
			\end{figure}
			
			As with the proper orthogonal modes, many of the dynamic modes can be associated with filling dynamics. While the dynamic modes generally come in complex conjugate pairs, the first dynamic modes are all purely real and again are visually indistinguishable from the time-averaged velocity fields in Fig.~\ref{fig:TimeAvg}. The first pair of dynamic modes (i.e., modes $2$ and $3$) captures the downstream progression of the mitral and regurgitant jets quite clearly in the real part (see Fig.~\ref{fig:DMDM002Re}), while the imaginary part bares some resemblance to the second POD mode (compare Figs.~\ref{fig:PODM002} and~\ref{fig:DMDM002Im}). Additionally, we observe some resemblance between several other proper orthogonal modes and the imaginary parts of the remaining dominant dynamic modes. The two decompositions for these flows therefore contain much of the same information in their respective most dominant modal structures, namely, in terms of energy for POD and in terms of penalized amplitude for DMD.
			
\begin{figure}[!t]
				\centering
				\includegraphics{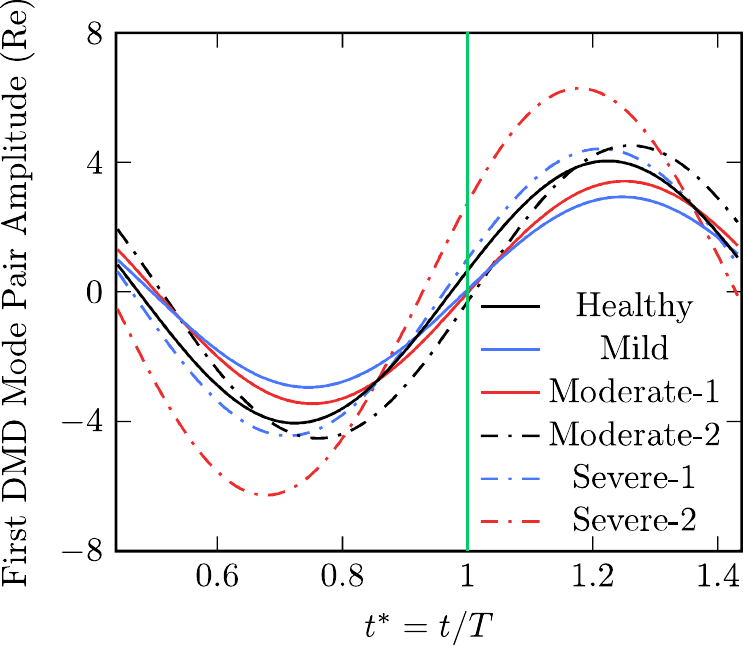}
				\caption{Real part of the temporal dynamics associated with the first pair of dynamic modes (i.e., modes $2$ and $3$). The frequency of the oscillations differs from the heart rate ($1.17$~Hz) by less than $0.3$\%. Note the phase shift of each case relative to the healthy scenario. The vertical green line at $t^* = 1$ marks the beginning of the ejection phase.}
				\label{fig:DMDTD}
			\end{figure}

			The real part of the temporal dynamics for the first mode pair is shown in Fig.~\ref{fig:DMDTD} for all cases; the imaginary part is of course a simple $90^{\circ}$ rightward phase shift of the real part. The first mode pair oscillates distinctly close to the heart rate of the simulations ($1.17$~Hz) for all cases, differing by less than $0.3$\%. The remaining mode pairs oscillate close to the harmonics of the heart rate, differing by less than $0.4$\% for all harmonics. In the case of DMD, the temporal dynamics in Fig.~\ref{fig:DMDTD} capture phase information that was not available when using POD in Fig.~\ref{fig:PODamps}. Particularly, in the work of \citeA{DiLabbioKadem18} and \citeA{DiLabbioVetelKadem18}, it was shown that the regurgitant inflow appeared to arrive in the field of view later than the mitral inflow for the mild ($\mathrm{ROA} = 3.3$\%) and moderate-$1$ ($\mathrm{ROA} = 5.9$\%) cases. By contrast, the regurgitant inflow arrived distinctly earlier for the severe cases ($\mathrm{ROA} = 17.2$ and $26.1$\%). This behavior is captured by the temporal dynamics of the first mode pair, where a rightward shift (delay) can be seen for the mild and moderate-$1$ cases and a leftward shift can be seen for the severe cases. This is to be contrasted with what was observed for POD, where the temporal dynamics of the first mode in Fig.~\ref{fig:PODamp1} show no distinct shift, while those of the second mode in Fig.~\ref{fig:PODamp2} seemingly show the opposite tendency if anything.
			
\begin{figure}[!b]
				\centering
				\subfloat[\label{fig:DMDenergy}]{%
					\includegraphics{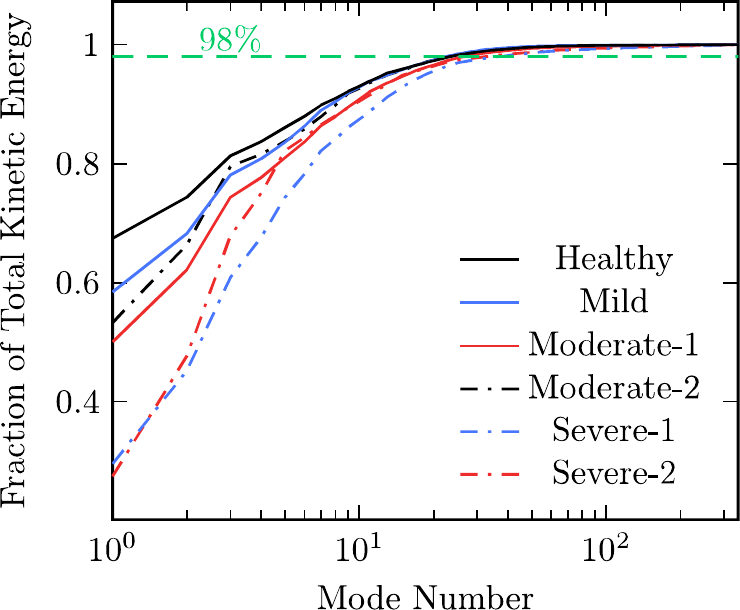}
				}\hfill
				\subfloat[\label{fig:DMDentropy}]{%
					\includegraphics{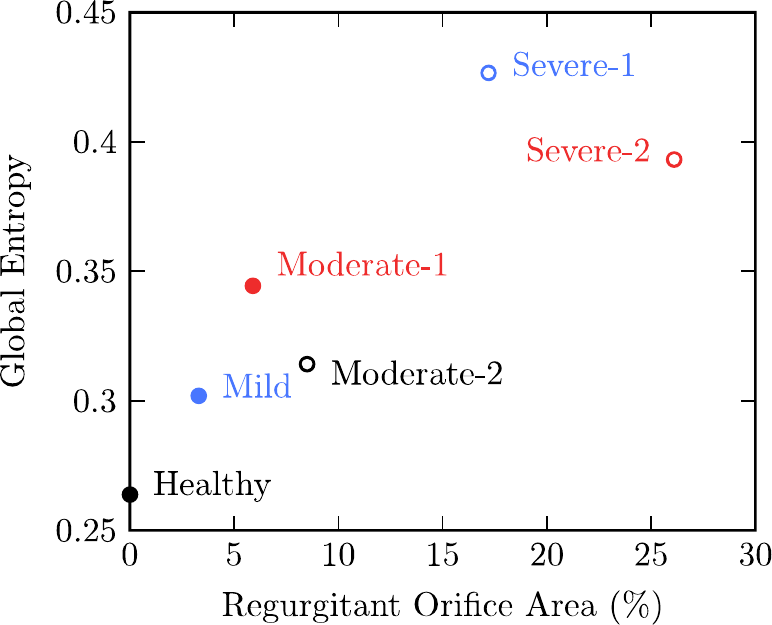}
				}
				\caption{(a) The fraction of accumulated kinetic energy against the mode number and (b) the corresponding Shannon entropy are shown for the dynamic mode decompositions of each case.}
				\label{fig:DMDchars}
			\end{figure}

			With regard to accumulation of energy, the DMD requires $129$, $115$, $225$, $163$, $293$, and $286$ modes, in order of increasing regurgitation severity, to reconstruct the flow fields and capture $99.9$\% of their ensemble kinetic energy; refer to Tab.~\ref{tab:DMDchars} in Sec.~\hyperref[sec:DMDperf]{\ref*{sec:DMD}.\ref*{sec:DMDperf}} for the number of modes required to capture $98.0$\%, $99.0$\%, and $99.5$\%. Additionally, the first dynamic modes capture $67.4$\%, $58.4$\%, $50.1$\%, $53.3$\%, $29.6$\%, and $27.5$\% of the respective ensemble flow kinetic energy, considerably less than the first proper orthogonal modes for each case, despite the modal structures being all effectively identical to the time-averaged flow fields for both POD and DMD; the difference of course arises from their respective temporal dynamics. This tendency is of course expected since, by construction, POD produces modal structures each having the largest possible kinetic energy in descending order. By contrast, DMD produces modal structures that can only contribute to the energy at fixed frequencies. Therefore, this highlights the temporal complexity of the kinetic energy in the intraventricular flows, suggesting that many frequencies are involved. The corresponding accumulation of energy with mode number and the associated Shannon entropy are shown in Figs.~\ref{fig:DMDenergy} and~\ref{fig:DMDentropy}, respectively. Note the much slower accumulation of kinetic energy for DMD as opposed to what was seen for POD in Fig.~\ref{fig:PODenergy} as well as the wider spread of energy among the modes suggested by the larger entropy values.
			
		\subsection{\label{sec:DMDperf}Performance of the DMD models}
			
			The effectiveness of the DMD models in reproducing the integral of the circulation per unit area and the viscous energy loss, as given by Eqs.~(\ref{eq:Circ}) and~(\ref{eq:VEL}), respectively, was computed and is shown in Tab.~\ref{tab:DMDchars}. Again, the general behavior of vorticity seems to be rather well-preserved by all the DMD models. The errors are rather acceptable for most cases when reconstructed to $98.0$\% of their respective ensemble flow kinetic energy, as was also observed for the POD models, except for the moderate-$1$ ($\mathrm{ROA} = 5.9$\%) and severe-$1$ ($\mathrm{ROA} = 17.2$\%) cases. Overall, the POD and DMD models produce similar errors for the circulation integral at any given energy level, although the POD models perform considerably better for the moderate-$1$ and severe-$1$ cases. With regard to the total energy dissipated by viscous stresses over one cardiac cycle, the monotonic increase in the reconstructed flow fields is again preserved for the DMD models with diminished magnitudes, with the linearity being better-preserved beginning with the $99.5$\% DMD model. The reconstruction errors of the POD and DMD models capturing $98.0$\% and $99.0$\% of the ensemble flow kinetic energy are comparable. However, as more kinetic energy is captured by the reconstructions, the POD  models outperform those of the DMD models in reproducing the viscous energy loss, particularly for the two severe cases ($\mathrm{ROA} = 17.2$ and $26.1$\%).
			
			To assess the ability of the DMD models to reproduce global advection behavior, we again consider the fraction of initial particles remaining in the left ventricle after releasing $\sim800\, 000$ particles at the start of the ejection phase and advecting them for four cardiac cycles. By the end of the first ejection phase ($t^* = 0.438$), the fraction of particles remaining exhibits an error of less than $0.4$\% for the DMD models capturing $98.0$\% of the respective ensemble flow kinetic energy for all cases. At an energy level of $99.0$\%, the errors fall below $0.22$\% for all cases, and at $99.9$\%, the errors fall below $0.02$\%. When considering two cardiac cycles, at an energy level of $99.5$\%, the models produce errors of less than $4$\%, whereas at an energy level of $99.9$\%, the errors fall to less than $0.6$\%. Up to four cardiac cycles, at an energy level of $99.9$\%, the models produce errors of less than $4$\%. At any given energy level, reconstruction from the DMD models reproduces the global particle advection patterns more accurately than do the POD models. While this may be a direct consequence of the DMD making use of more modes, it does not explain why the errors for the viscous energy loss are generally on par or poorer for DMD than for POD. Rather, the flow is dependent on many frequencies to dissipate energy (the small-scale flow), while the advection of particles is mostly governed by lower frequency dynamics (the large-scale flow). Consequently, with the dominant dynamic modes concentrated in the low frequency regime, characteristics such as the viscous dissipation are more difficult to capture, while global particle advection patterns are easily reconstructed. By contrast, the dominant proper orthogonal modes are composed of many frequencies, and so while they perform better at capturing viscous dissipation, these high frequencies may pollute the large-scale flow that governs particle advection.
			
			\begin{table}[!h]
				\begin{center}
					\begin{tabular}{r*{15}{c}}
						\multicolumn{15}{l}{Table \ref*{tab:DMDchars}: DMD reconstruction characteristics and errors} \\
						\hline\hline\\[-1.0em]
						& \multicolumn{4}{c}{Number of Modes} & & \multicolumn{4}{c}{Circulation Error (\%)} & & \multicolumn{4}{c}{Energy Loss Error (\%)} \\
						\cline{2-5}
						\cline{7-10}
						\cline{12-15}\\[-1.0em]
						Captured KE (\%) $\rightarrow$ & $98.0$ & $99.0$ & $99.5$ & $99.9$ & & $98.0$ & $99.0$ & $99.5$ & $99.9$ & & $98.0$ & $99.0$ & $99.5$ & $99.9$ \\
						\hline\\[-1.0em]
						Healthy (\phantom{$0$}$0$\phantom{$.0$}\% $\mathrm{ROA}$) & $23$ & $35$ & $49$ & $129$ & & $2.15$ & $1.85$ & $1.23$ & $0.44$ & & $21.7$ & $16.8$ & $13.6$ & $6.71$ \\
						Mild (\phantom{$0$}$3.3$\% $\mathrm{ROA}$) & $23$ & $31$ & $41$ & $115$ & & $1.43$ & $1.10$ & $0.86$ & $0.31$ & & $17.2$ & $13.0$ & $9.73$ & $3.55$ \\
						Moderate-1 (\phantom{$0$}$5.9$\% $\mathrm{ROA}$) & $27$ & $41$ & $61$ & $225$ & & $22.9$ & $17.6$ & $10.6$ & $1.90$ & & $20.1$ & $15.2$ & $11.8$ & $3.15$ \\
						Moderate-2 (\phantom{$0$}$8.5$\% $\mathrm{ROA}$) & $23$ & $33$ & $47$ & $163$ & & $2.29$ & $1.92$ & $1.36$ & $0.37$ & & $22.7$ & $18.0$ & $14.1$ & $5.11$ \\
						Severe-1 ($17.2$\% $\mathrm{ROA}$) & $39$ & $67$ & $133$ & $293$ & & $19.7$ & $9.77$ & $3.44$ & $0.37$ & & $34.6$ & $27.9$ & $18.6$ & $4.14$ \\
						Severe-2 ($26.1$\% $\mathrm{ROA}$) & $33$ & $61$ & $119$ & $286$ & & $3.19$ & $1.83$ & $0.89$ & $0.12$ & & $38.6$ & $31.1$ & $22.4$ & $6.69$ \\
						\hline\hline
					\end{tabular}
				\end{center}
				\refstepcounter{table}\label{tab:DMDchars}
			\end{table}
			
			Comprehensively, DMD produces reduced-order models of comparable accuracy to those obtained using POD. The ability of the DMD models to capture blood stasis behavior over that of the POD models is however considerably improved, although at the expense of requiring more modes for a given energy level. With the $99.5$\% DMD model outperforming that for POD for the healthy intraventricular flow using only $20$ additional modes, it is this reduced-order model that is made available on \href{https://github.com/dilabbiog/ROMs--LV_flow_with_AR}{GitHub} and discussed in the \href{ftp://ftp.aip.org/epaps/phys_fluids/E-PHFLE6-31-006903}{supplementary material}. The same decision was made for the mild ($\mathrm{ROA} = 3.3$\%) and moderate ($\mathrm{ROA} = 5.9$ and $8.5$\%) cases of aortic regurgitation. The $99.5$\% DMD models for the severe cases ($\mathrm{ROA} = 17.2$ and $26.1$\%) however require far too many more modes without any significant improvement over what was obtained for the $99.5$\% POD models. Therefore, the $99.5$\% POD models for the severe cases of regurgitation are made available on \href{https://github.com/dilabbiog/ROMs--LV_flow_with_AR}{GitHub} and discussed in the \href{ftp://ftp.aip.org/epaps/phys_fluids/E-PHFLE6-31-006903}{supplementary material} rather than those for DMD.
			
	\section{\label{sec:Conc}Conclusion}
		
		Reduced-order models for \textit{in vitro} datasets of the intraventricular flow for a healthy left ventricle as well as for five severities of aortic regurgitation were constructed using POD and DMD. The models were based on the number of modes required to reconstruct the datasets and capture $98.0$\%, $99.0$\%, $99.5$\%, and $99.9$\% of their ensemble flow kinetic energy. The performance of the models in reconstructing the acquired flow fields was investigated by evaluating their ability to reproduce circulation, viscous energy dissipation, and global particle advection behavior. The selected models are made available on \href{https://github.com/dilabbiog/ROMs--LV_flow_with_AR}{GitHub} with additional details begin provided in the \href{ftp://ftp.aip.org/epaps/phys_fluids/E-PHFLE6-31-006903}{supplementary material}. The results of this work additionally emphasize several important conclusions. ($1$) At any given energy level, it appears that both POD and DMD preserve velocity gradients to similar accuracy based on the reconstruction errors obtained for total flow circulation and viscous energy loss. ($2$) Dynamic mode decomposition however requires more modes to achieve each energy level, highlighting the temporal complexity of the intraventricular flows (i.e., the viscous dissipation, and therefore the kinetic energy, relies on small-scale dynamics and so requires many frequencies to be modeled). ($3$) In all cases, DMD was observed to perform better at preserving global particle advection behavior using fewer modes, suggesting that the observed behavior tends to be less complex in time and hence is better-reconstructed using DMD (i.e., the advection is governed by large-scale dynamics and so requires fewer frequencies to be modeled). ($4$) This study additionally highlights a key point associated with the application of DMD to periodic velocity data acquired from PIV. The sensitivity of DMD to periodic experimental data, namely, to the correspondence between the first and final state vectors used in the decomposition, is an important factor to consider when generating reduced-order models or when using DMD for smoothing/filtering of the acquired velocity fields. Periodic flows should be governed by temporal dynamics that vary as pure sinusoids with little growth/decay, suggesting that the discrete-time eigenvalues correspond to roots of unity or, equivalently, that the continuous-time eigenvalues have a negligible real part. Given a complete period, by shifting the order of the snapshots as in Eq.~(\ref{eq:ordering}), the resulting temporal dynamics may better-represent the flow in question (as was the case in this study). ($5$) This study represents a first step in constructing a database of regurgitant intraventricular flows that may ultimately be used to predict the flows associated with any given regurgitant orifice area, similar to the methodology used in \citeA{McGregor08, McGregor09}, \citeA{McLeod10} and \citeA{Guibert14}. Currently, the number of cases included in this study is not sufficient to construct such a predictive model. Nonetheless, with such a model, assessment of the regurgitant orifice area in practice could then be associated with a complete intraventricular flow deduced without simulation, making flow-field-oriented clinical parameters readily computable in practice.
		
		While this study provides data-driven flow models for both healthy left ventricular flow as well as those subject to varying grades of aortic regurgitation, it must be understood that the models themselves come with several important limitations. First and foremost, the reduced-order models are here provided for a single two-dimensional plane of what are otherwise three-dimensional flows. While the plane of choice has an important practical value, being widely used by clinicians, a three-dimensional description would ultimately be required to capture the full range of phenomena. However, in three dimensions, the intraventricular flows will be inherently more complex and will certainly require additional modes to be adequately modeled. For instance, for a healthy left ventricular flow, the mitral inflow produces a full vortex ring, which is advected and deformed in three dimensions with time while part of it dissipates against the ventricle wall \cite{PedrizzettiDomenichini15}. In the plane of symmetry considered throughout this work, this behavior is not captured as the flow simply retains a two-dimensional swirl with a core of varying strength and position \cite{DiLabbioKadem18, DiLabbioVetelKadem18}. Furthermore, with increasing regurgitation severity, the regurgitant jet has been previously shown to be accompanied by turbulence \cite{DiLabbioVetelKadem18}. This of course occurs in three dimensions, greatly increasing the complexity of the flow, particularly surrounding the jet itself. It is therefore likely that the viscous dissipation characteristics in the case of aortic regurgitation will be further underestimated by a reduced-order model. Nonetheless, while this group is working toward such a three-dimensional description, there is currently a lack of three-dimensional \textit{in vivo} flow data with which to validate the results in the case of aortic regurgitation. Further limitations specifically regarding the experimental conditions can be found in the work of \citeA{DiLabbioKadem18} and more so in the work of \citeA{DiLabbioVetelKadem18}. An additional limitation, inherent to the reduced-order modeling techniques, is that we have here used methods effectively decomposing the flow into modes based on the energy or frequency content of the velocity field, the underlying assumption being that the flows are well-represented in such a context. This raises the question as to whether intraventricular flows are better described by some other framework, such as using vorticity or strain rate. In the former, a POD would rank the modes according to an approximation of their ensemble enstrophy and in the latter of their ensemble viscous energy dissipation (with the exception of a multiplicative constant). More generally, perhaps the full Cauchy-Stokes description would better-represent such flows, or even Lagrangian descriptions such as the finite-time Lyapunov exponent. If this should be the case, a model reduction technique ought to reproduce the flows from an even further restricted set of modes.
		
	\section*{Supplementary Material}
		
		The details regarding how the models were generated, how to reconstruct the flows from the data, and the errors associated with the reconstructed flows are provided in the \href{ftp://ftp.aip.org/epaps/phys_fluids/E-PHFLE6-31-006903}{supplementary material}. The associated MATLAB scripts are also provided which may be used as pseudocode for other programming languages. The reduced-order models discussed in the \href{ftp://ftp.aip.org/epaps/phys_fluids/E-PHFLE6-31-006903}{supplementary material} capture $99.5$\% of the ensemble flow kinetic energy. For the healthy intraventricular flow as well as the mild and moderate cases of aortic regurgitation, the models were produced using dynamic mode decomposition while for the severe cases, the models were produced using proper orthogonal decomposition (refer to the main text for further discussion regarding this decision).
		
	\section*{Acknowledgments}
		
		This work was supported by a grant from the Natural Sciences and Engineering Research Council of Canada (Grant No.\ $343164$-$07$). G.D.L.\ was supported by the Vanier Canada Graduate Scholarship. The authors are indebted to the reviewers for their excellent suggestions which have greatly improved the clarity and impact of this work. The authors would also like to thank Team $16$ of the $2013$-$14$ Capstone design project at Concordia University (Montr\'{e}al, QC) for their hard work in the development of the heart simulator from which the data have been acquired, particularly Alexandre B\'{e}langer, Emilia Benevento, and Nick Ghaffari as well as Yves-Christian Tchatchouang for further improvements made to the system.
		
	\appendix
	\renewcommand{\theequation}{\thesection.\arabic{equation}}
	\renewcommand{\thefigure}{\thesection.\arabic{figure}}
	\renewcommand{\thetable}{\thesection.\arabic{table}}
	
	\section{\label{app:PODAlg}Proper Orthogonal Decomposition}
	\setcounter{equation}{0}
	\setcounter{figure}{0}
	\setcounter{table}{0}
		
		At its core, POD looks to spatially decorrelate the velocity signals at points throughout a flow over time. The method begins by arranging the sequence of $n$ velocity fields, or flow snapshots, each having $m$ grid points with velocity components $u$ and $v$, in a tall matrix, i.e.,
		\begin{equation}
		\label{eq:stateVars}
			\mathbf{X} = 
			\begin{bmatrix}
				\mid         &        & \mid            \\
				\mathbf{x}_1 & \cdots & \mathbf{x}_n    \\
				\mid         &        & \mid
			\end{bmatrix}
			\qquad \text{where} \qquad
			\mathbf{x}_k = 
			\begin{bmatrix}
				u_{1k} \\ \vdots \\ u_{mk} \\ v_{1k} \\ \vdots \\ v_{mk}
			\end{bmatrix}.
		\end{equation}
		In this work, we follow the snapshot method of \citeA{Sirovich87a} and compute the eigenvectors $\mathbf{q}_k$ of the temporal correlation matrix $\mathbf{C} = \mathbf{X}^{\mathrm{T}}\mathbf{X}$ from
		\begin{equation}
		\label{eq:PODdiag}
			\mathbf{C}\mathbf{Q} = \mathbf{Q}\boldsymbol{\Lambda}.
		\end{equation}
		After sorting the eigenvalues $\lambda_k$ and corresponding eigenvectors $\mathbf{q}_k$ in descending order, the proper orthogonal modes $\boldsymbol\phi_k$ are then given by the columns of the projection
		\begin{equation}
		\label{eq:PODmodes}
			\boldsymbol{\Phi} = \mathbf{X}\mathbf{Q}\boldsymbol{\Lambda}^{-1/2}.
		\end{equation}
		With the proper orthogonal modes representing a new basis, the flow at any given time $\mathbf{x}_k$ can be expressed as a linear combination of the modes, i.e.,
		\begin{equation}
		\label{eq:PODlincomb}
			\mathbf{x}_k = \sum_{l = 1}^n b_{kl}\boldsymbol{\phi}_l \qquad \text{or} \qquad \mathbf{X} = \boldsymbol{\Phi}\mathbf{B},
		\end{equation}
		where the amplitudes for reconstruction of the flow at time $k$ are given by the columns of
		\begin{equation}
		\label{eq:PODamps}
			\mathbf{B} = \boldsymbol{\Phi}^{\mathrm{T}}\mathbf{X}.
		\end{equation}
		Alternatively, the rows of $\mathbf{B}$, commonly referred to as the temporal modes or temporal dynamics, represent the amplitude signals of each mode over time.
		
	\section{\label{app:DMDAlg}Exact Dynamic Mode Decomposition}
	\setcounter{equation}{0}
	\setcounter{figure}{0}
	\setcounter{table}{0}
		
		In exact DMD, the dynamic modes are defined as the exact eigenvectors of the approximating linear operator between two sets of data $\mathbf{X}$ and $\mathbf{Y}$ \cite{Tu14}. Denoting this operator by $\mathbf{A}$, a linear relation between the data, $\mathbf{Y} \approx \mathbf{A}\mathbf{X}$, is solved for in the least squares sense, giving
		\begin{equation}
		\label{eq:DMDlinop}
			\mathbf{A} = \mathbf{Y}\mathbf{X}^+,
		\end{equation}
		where the superscript ``+'' denotes the Moore-Penrose pseudoinverse. In our case, we have a sequential time series and so $\mathbf{X}$ and $\mathbf{Y}$ are defined as
		\begin{equation}
		\label{eq:DMDdata}
			\mathbf{X} =
			\begin{bmatrix}
				\mid         & \mid         &        & \mid             \\
				\mathbf{x}_1 & \mathbf{x}_2 & \cdots & \mathbf{x}_{n-1} \\
				\mid         & \mid         &        & \mid
			\end{bmatrix} \qquad \text{and} \qquad \mathbf{Y} = 
			\begin{bmatrix}
				\mid         & \mid         &        & \mid           \\
				\mathbf{x}_2 & \mathbf{x}_3 & \cdots & \mathbf{x}_{n} \\
				\mid         & \mid         &        & \mid
			\end{bmatrix}.
		\end{equation}
		As proposed by \citeA{Schmid10}, given the reduced singular value decomposition of $\mathbf{X} = \mathbf{U}\boldsymbol{\Sigma}\mathbf{V}^*$ (the superscript ``$*$'' denoting the complex conjugate transpose), the eigenvalues of $\mathbf{A}$ are simply given by the eigenvalue problem of the similar matrix $\widetilde{\mathbf{A}} = \mathbf{U}^*\mathbf{Y}\mathbf{V}\boldsymbol{\Sigma}^{-1}$, i.e.,
		\begin{equation}
		\label{eq:DMDeigvals}
			\widetilde{\mathbf{A}}\mathbf{W} = \mathbf{W}\boldsymbol{\Lambda}.
		\end{equation}
		Then, as proven by \citeA{Tu14}, the exact eigenvectors of $\mathbf{A}$, or the dynamic modes $\boldsymbol{\psi}_k$, are given by the columns of
		\begin{equation}
		\label{eq:DMDeigvecs}
			\boldsymbol{\Psi} = \mathbf{Y}\mathbf{V}\boldsymbol{\Sigma}^{-1}\mathbf{W}.
		\end{equation}
		In this work, we follow \citeA{Tu14} and normalize the dynamic modes $\boldsymbol\psi_k$ in Eq.~(\ref{eq:DMDeigvecs}) by their eigenvalues, giving
		\begin{equation}
		\label{eq:DMDmodes}
			\boldsymbol{\Psi} = \mathbf{Y}\mathbf{V}\boldsymbol{\Sigma}^{-1}\mathbf{W}\boldsymbol{\Lambda}^{-1}.
		\end{equation}
		While this is not necessary, it does have an interesting consequence in this work in relation to the optimal amplitudes $\left(\alpha_{\mathrm{dmd}}\right)$ derived by \citeA{Jovanovic14} for standard DMD, which we will discuss shortly. The temporal information of the dynamic modes are effectively given by the discrete-time eigenvalues (or Ritz values) $\lambda_k$, which are in general complex, having magnitude $|\lambda_k|$ representing the growth or decay rate of the dynamic mode $\boldsymbol{\psi}_k$ and principal argument $\mathrm{Arg}\left(\lambda_k\right)$ representing its phase. In the complex plane, an eigenvalue $\lambda_k$ points to the instability of mode $k$ if it lies outside the unit circle. In order to investigate stability in the classical sense, where a positive real part denotes unstable exponential growth, the logarithmic mapping $\gamma_k = \ln\left(\lambda_k\right)/\Delta{t}$ must be performed. In this case, the real part $\sigma_k = \mathrm{Re}\left(\gamma_k\right)$ represents the exponential growth or decay rate and the imaginary part $\omega_k = \mathrm{Im}\left(\gamma_k\right)$ represents the radial frequency. In the case of a sequential time series with uniformly spaced samples in time ($\Delta t$), a fit to the original dataset is given by
		\begin{equation}
		\label{eq:DMDreconstruct}
			\mathbf{x}_{k+1} \approx \boldsymbol{\Psi}\boldsymbol{\Lambda}^k\mathbf{b} \qquad \text{or} \qquad \mathbf{x}_{k+1} \approx \boldsymbol{\Psi}e^{\left(k\Delta t\right)\boldsymbol{\Gamma}}\mathbf{b},
		\end{equation}
		where $\mathbf{b}$ is a set of coefficients or modal amplitudes, $\boldsymbol\Gamma$ is a diagonal matrix containing the elements of $\gamma_k$, and $k = 0, 1, ..., n-1$. Although $\mathbf{b}$ may be obtained by solving Eq.~(\ref{eq:DMDreconstruct}) with $k = 0$, following \citeA{Tu14}, we opted to use $k = 1$, giving
		\begin{equation}
		\label{eq:DMDamps}
			\mathbf{b} = \boldsymbol{\Lambda}^{-1}\boldsymbol{\Psi}^+\mathbf{x}_2.
		\end{equation}
		The choice is based on the fact that $\mathbf{x}_2$ ought to be in the range of $\mathbf{A}$ (i.e., $\mathbf{x}_1$ does not necessarily lie in the column space of $\mathbf{Y}$), and so these approximating coefficients should better represent the data. Using Eq.~(\ref{eq:DMDmodes}) for the modes and Eq.~(\ref{eq:DMDamps}) for the amplitudes, as was performed throughout Sec.~\ref{sec:DMD}, the resulting amplitudes differ from the optimal amplitudes $\left(\alpha_{\mathrm{dmd}}\right)$ on the order of $10^{-12}$ to $10^{-14}$ for the datasets used in this work. By contrast, using Eq.~(\ref{eq:DMDeigvecs}) for the modes and Eq.~(\ref{eq:DMDamps}) for the amplitudes, or making use of the relation $\mathbf{b} = \boldsymbol{\Psi}^+\mathbf{x}_1$, there is a difference on the order of $10^{-2}$ to $10^{-4}$.
		
		With the amplitudes obtained, if the Vandermonde matrix of the eigenvalues is defined as
		\begin{equation}
		\label{eq:Vand}
			\mathbf{T} =
			\begin{bmatrix}
				1      & \lambda_1      & \lambda_1^2     & \cdots & \lambda_1^{n-2}\\
				1      & \lambda_2      & \lambda_2^2     & \cdots & \lambda_2^{n-2}\\
				\vdots & \vdots         & \vdots          & \ddots & \vdots         \\
				1      & \lambda_{n-1}  & \lambda_{n-1}^2 & \cdots & \lambda_{n-1}^{n-2}\\
			\end{bmatrix},
		\end{equation}
		the approximation of the dataset is given compactly as
		\begin{equation}
		\label{eq:DMDapprox}
			\mathbf{X} \approx \boldsymbol{\Psi}\mathbf{B}\mathbf{T},
		\end{equation}
		where $\mathbf{B}$ is a diagonal matrix containing the elements of $\mathbf{b}$. The product $\mathbf{B}\mathbf{T}$ describes the temporal dynamics of the modes.
		
	\section{\label{app:EnsAvg}Effect of Ensemble-Averaging Step on Viscous Energy Dissipation}
	\setcounter{equation}{0}
	\setcounter{figure}{0}
	\setcounter{table}{0}
		
		In this work, each reduced-order model is constructed from the ensemble-average of the ten time-resolved acquisitions made for each respective case. The results presented in our previous studies are largely based on using the ensemble-averaged flows except in cases where turbulent fluctuations become important, such as in our investigation of the rate of energy dissipated by viscous stresses \cite{DiLabbioKadem18} and of the time-frequency spectra of selected velocity signals \cite{DiLabbioVetelKadem18}. This being the case, and given that many of the underlying phenomena have been described in our previous studies, it was decided that constructing a reduced-order model of the ensemble-averaged flows would be more beneficial to researchers interested in using the provided data. Additionally, it should be understood that data-driven reduced-order modeling techniques would nonetheless smooth the data if applied to any one of the ten time-resolved acquisitions, and so using the ensemble-averaged flows would effectively produce the same smooth results with improved convergence of the models. However, use of the ensemble-average does come with the important consequence that many turbulent fluctuations will be inherently filtered out. Nonetheless, as we demonstrate in Secs.~\ref{sec:POD} and~\ref{sec:DMD}, many of the reported results can still be effectively reproduced, with the exception of the magnitudes of viscous dissipation and the highest frequencies appearing in the time-frequency spectra. We do note, however, that the monotonically increasing trend in viscous energy loss is still preserved for the regurgitant cases using the ensemble-averaged flows. Furthermore, the corresponding error in viscous energy loss was found to be lowest for the healthy scenario. Therefore, the reduced-order model constructed particularly for healthy left ventricular flow ought to more accurately represent what has been observed in the literature, which is most desirable to the general reader given its broader application.
		
		\begin{figure}[!t]
			\centering
			\subfloat[\label{fig:VEDrunAvg}]{%
				\includegraphics{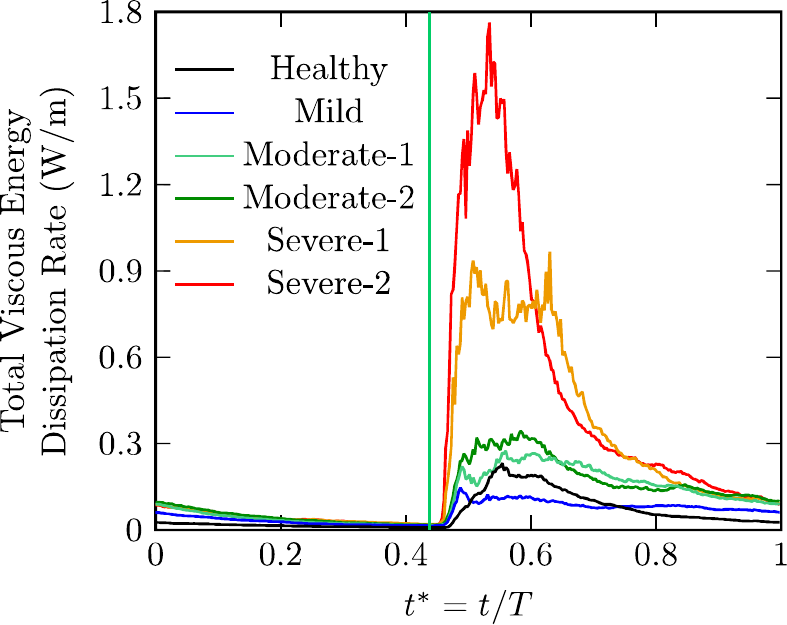}
			}\hfill
			\subfloat[\label{fig:VEDensAvg}]{%
				\includegraphics{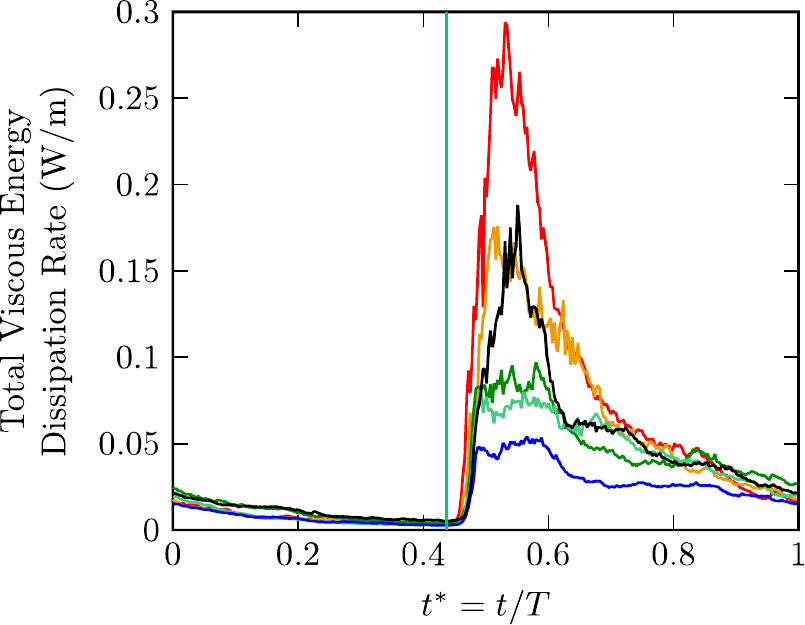}
			}
			\caption{In (a), the total rate of energy dissipation by viscous stresses was computed for each of the ten time-resolved acquisitions and then ensemble-averaged, whereas in (b), it was computed for the ensemble-averaged flows. The vertical green line at $t^* = 0.438$ marks the beginning of the filling phase. (a) is reprinted with permission from G.\ Di Labbio and L.\ Kadem, ``Jet collisions and vortex reversal in the human left ventricle,'' \href{https://doi.org/10.1016/j.jbiomech.2018.07.023}{J.\ Biomech.}~78, 155-160 (2018). Copyright 2018 Elsevier.}
			\label{fig:VEDchars}
		\end{figure}
		
		\begin{figure}[!b]
			\centering
			\subfloat[\label{fig:VEL}]{%
				\includegraphics{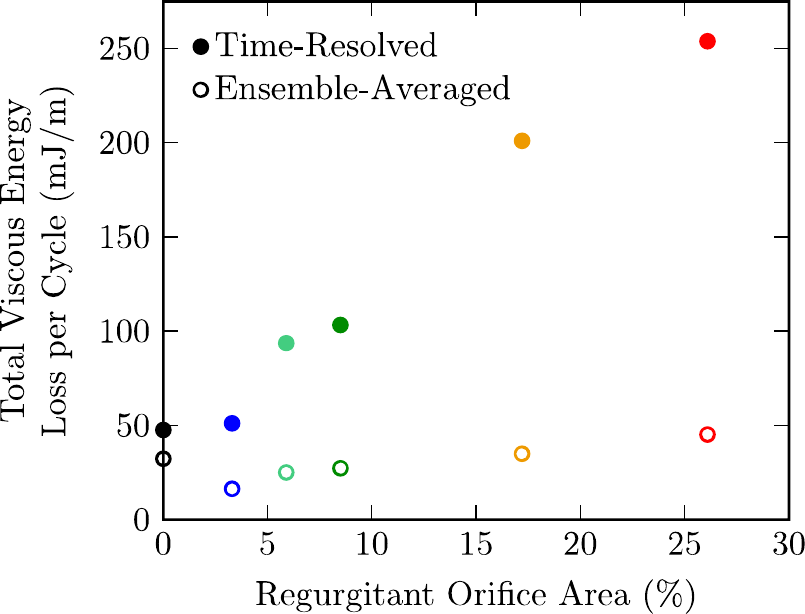}
			}\hfill
			\subfloat[\label{fig:VELerr}]{%
				\includegraphics{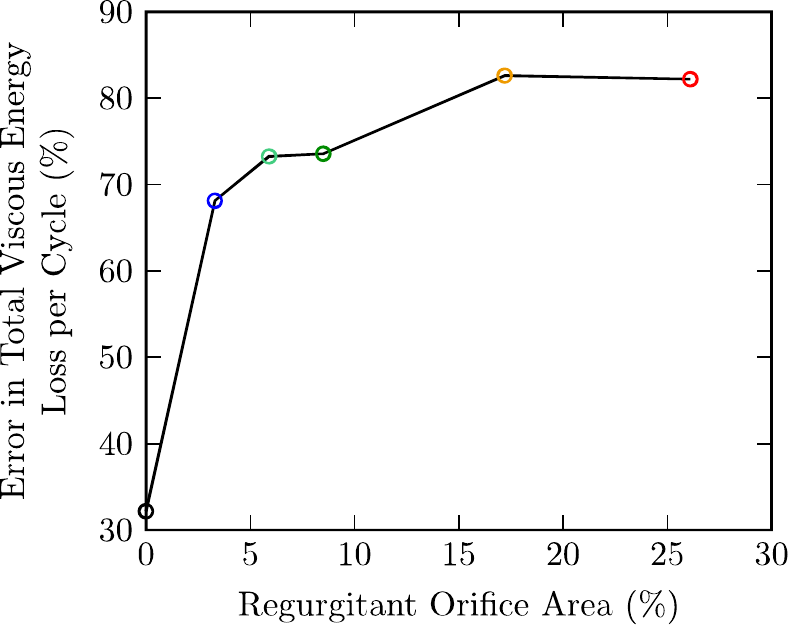}
			}
			\caption{In (a), the total viscous energy loss per cardiac cycle (per unit depth) was computed for each of the ten time-resolved acquisitions and then ensemble-averaged (filled circles). Additionally, the calculation was performed for the ensemble-averaged flows (open circles). The corresponding error in the viscous energy loss for the ensemble-averaged flows is shown in (b).}
			\label{fig:VELchars}
		\end{figure}
		
		We have previously shown that the energy dissipated by viscous stresses per cardiac cycle appears to increase monotonically with regurgitation severity for the acquired datasets \cite{DiLabbioKadem18}. Much of these stresses are however associated with the turbulent propagation of the regurgitant jet, a feature illustrated further in the work of \citeA{DiLabbioVetelKadem18}. With these turbulent velocity fluctuations contributing significantly to the dissipation of energy, the ensemble-averaging results in an appreciably poor underestimate of the rate of energy dissipation in all simulated cases. This has important repercussions for acquisition techniques that inherently rely on phase-averaging such as magnetic resonance imaging or phase-locked particle image velocimetry. For comparison, Fig.~\ref{fig:VEDrunAvg} shows the total rate of viscous dissipation throughout the cardiac cycle for all cases as reported in the work of \citeA{DiLabbioKadem18} and Fig.~\ref{fig:VEDensAvg} shows the result computed on the ensemble-averaged flows. Not only is the drop in magnitude evident, but worse, it appears that the healthy scenario dissipates almost as much energy as the severe-$1$ case ($\mathrm{ROA} = 17.2$\%). Although unfortunate, the resulting misrepresentation of the monotonically increasing trend of energy dissipation should be expected. The flow in the healthy scenario is largely laminar and so naturally possesses lower magnitude velocity fluctuations. Additionally, the energy dissipated in the healthy scenario arises mainly from the shear layer of the mitral inflow. This being the case, the healthy intraventricular flow is better-represented by the ensemble-average than the regurgitant cases, and so its energy dissipation profile in Fig. 14(a) resembles that in Fig.~\ref{fig:VEDrunAvg} more closely. On the contrary, with aortic regurgitation, the mitral inflow contributes very little to the total energy dissipation in comparison with the disturbances caused by the regurgitant jet. In order to evaluate the corresponding error, the integrals of the energy dissipation curves over one cardiac cycle are computed and shown in Fig.~\ref{fig:VEL}, while the corresponding
		errors are shown in Fig.~\ref{fig:VELerr}. The errors increase consistently with regurgitation severity, ranging from $32.2$\% for the healthy intraventricular flow to $82.6$\% in the second most severe case of aortic regurgitation ($\mathrm{ROA} = 17.2$\%). Nevertheless, the monotonic increase does still present itself when considering only the diseased cases, likely due entirely to the increasing shear observed in the shear layer of the regurgitant jet with increasing severity.
		
	\bibliographystyle{apacite}

\end{document}